\documentclass[aps,prl,reprint,groupedaddress]{revtex4-1}

\usepackage[dvips]{graphicx}
\usepackage{amsmath,amsthm,array} 
\usepackage{multirow}
\usepackage{color}
\usepackage{bm}
\usepackage{lipsum} 
\usepackage{ulem} 
\usepackage{multirow}
\usepackage{array}
\newcolumntype{L}[1]{>{\raggedright\let\newline\\\arraybackslash\hspace{0pt}}m{#1}}
\newcolumntype{C}[1]{>{\centering\let\newline\\\arraybackslash\hspace{0pt}}m{#1}}
\newcolumntype{R}[1]{>{\raggedleft\let\newline\\\arraybackslash\hspace{0pt}}m{#1}}
\usepackage{algorithm}
\usepackage{algorithmic}
\usepackage{enumitem}
\setlist[enumerate]{itemsep=0mm}
\usepackage{cleveref}

\usepackage{mdframed}    
\newmdenv[skipabove=6mm, skipbelow=6mm]{kotak}   
\DeclareMathOperator*{\argmax}{arg\,max}

\usepackage{soul}

\begin{document}

\title{Explainable Machine Learning for Materials Discovery: Predicting the Potentially Formable Nd-Fe-B Crystal Structures and Extracting Structure-Stability Relationship}

\author{
Tien-Lam Pham$^{1,3}$, Duong-Nguyen Nguyen$^{1}$, Minh-Quyet Ha$^{1}$, Hiori Kino$^{2,4}$, Takashi Miyake$^{2,3,4}$, Hieu-Chi Dam$^{1,2,5}$}
\affiliation{
$^{1}$Japan Advanced Institute of Science and Technology, 1-1 Asahidai, Nomi, Ishikawa 923-1292, Japan\\
$^{2}$Center for Materials Research by Information Integration, National Institute for Materials Science, 1-2-1 Sengen, Tsukuba, Ibaraki 305-0047, Japan\\
$^{3}$ESICMM, National Institute for Materials Science 1-2-1 Sengen, Tsukuba, Ibaraki 305-0047, Japan\\
$^{4}$CD-FMat, AIST, 1-1-1 Umezono, Tsukuba, Ibaraki 305-8568, Japan\\
$^{5}$JST, PRESTO, 4-1-8 Honcho, Kawaguchi, Saitama 332-0012, Japan}

\date{\today}

\begin{abstract}
New Nd-Fe-B crystal structures can be formed via the elemental substitution of LA-T-X host structures, including lanthanides (LA), transition metals (T), and light elements X = B, C, N, and O. The 5967 samples of ternary LA-T-X materials that are collected are then used as the host structures. For each host crystal structure, a substituted crystal structure is created by substituting all lanthanide sites with Nd, all transition metal sites with Fe, and all light element sites with B. High-throughput first-principles calculations are applied to evaluate the phase stability of the newly created crystal structures, and 20 of them are found to be potentially formable. A data-driven approach based on supervised and unsupervised learning techniques is applied to estimate the stability and analyze the structure--stability relationship of the newly created Nd-Fe-B crystal structures. For predicting the stability for the newly created Nd-Fe-B structures, three supervised learning models, kernel ridge regression, logistic classification, and decision tree model, are learned from the LA-T-X host crystal structures; the models achieve the maximum accuracy and recall scores of $70.4\%$ and $68.7\%$, respectively. On the other hand, our proposed unsupervised learning model based on the integration of descriptor-relevance analysis and a Gaussian mixture model (GMM) achieves an accuracy and recall score of $72.9\%$ and $82.1\%$, respectively, which are significantly better than those of the supervised models. While capturing and interpreting the structure--stability relationship of the Nd-Fe-B crystal structures, the unsupervised learning model indicates that the average atomic coordination number and coordination number of the Fe sites are the most important factors in determining the phase stability of the new substituted Nd-Fe-B crystal structures.

\end{abstract}


\maketitle

\section{1. Introduction}\label{introduction}
The major challenge in finding new stable material structures in nature requires high-throughput screening of an enormous number of candidate structures, which are generated from different atomic arrangements in three-dimensional space. In fact, only a handful number of structures among these candidates are likely to exist. Therefore, for the non-serendipitous discovery of new materials, candidate structures must be generated strategically so that the screening space is reduced without overlooking potential materials.

A number of strategies have been proposed for the high-throughput screening processes \cite{Butler18, Curtarolo13, Saal2013} for finding various new materials. Almost all well-known screening methods consider first-principles calculations as the basis for the estimation of physical properties. Screening processes have been successfully developed for theoretically understanding rare-earth-lean intermetallic magnetic compounds \cite{Korner16, KORNER18}, Heusler compounds \cite{PhysRevB.95.024411, Jiangang18, Balluff17}, topological insulators \cite{Yang11, Li_2018}, perovskite materials \cite{Emery16, MICHALSKY2017124}, cathode coatings for Li-ion batteries \cite{Aykol16}, and $M_{2}AX$ compounds \cite{Ashton16}. In recent years, various screening processes have been used to replace canonical approaches by machine learning (ML) methods. A few notable works based on ML models involve searching for hard-magnetic phases \cite{MOLLER18}, Heusler compounds \cite{PhysRevMaterials.2.123801}, bimetallic facets catalysts \cite{Ulissi17}, BaTiO3-based piezoelectrics \cite{Xue13301}, polymer dielectrics \cite{Mannodi16}, perovskite halides \cite{Pilania16}, and low-thermal-hysteresis NiTi-based shape memory alloys \cite{Xue16}.

ML is expected to play three different roles in performing screening processes. The first role is to replace the density functional theory (DFT) calculation and reduce the calculation cost of physical property estimation, e.g., convex-hull distance \cite{PhysRevMaterials.2.123801} and adsorption energy \cite{Ulissi17}. The reported models have achieved reasonable results in statistical evaluation tests, such as cross validation. However, ensuring the reliability of extrapolating the physical properties of new materials is a major problem because the new screening materials do not always possess the same distribution as the training materials. 

\begin{figure*}
	\includegraphics[width=0.8\textwidth]{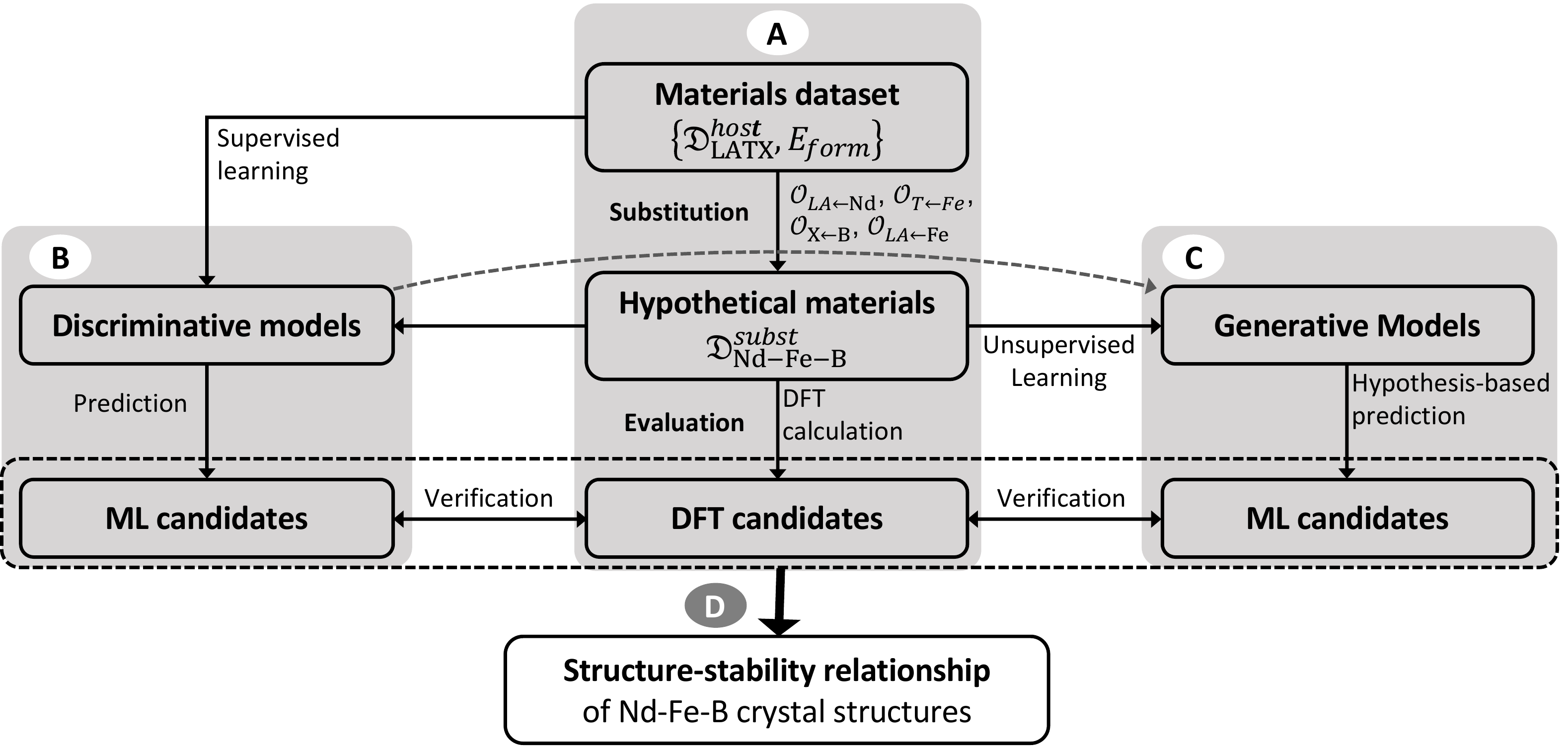}
    \caption{\label{fig:workflow} Workflow for extracting structure--stability relationship of Nd-Fe-B crystal structures by integrating high-throughput first-principles calculations, supervised learning, and unsupervised learning techniques.}
\end{figure*}

The second role of ML is to increase the success rate in screening processes. Given a list of hypothetical structures, ML methods are utilized for  recommending the most likely new potential materials using probabilistic models (e.g., Bayesian optimization techniques \cite{Yamashita18, Xue13301}). This approach requires a list of potential candidates to be prepared as an input, which is primarily based on human intuition. The bottleneck of the current recommendation methods is that a large number of known property materials are required as references for the system to start an effective recommendation process. This number increases dramatically with the material description dimension. Furthermore, the computational cost of the recommended process increases significantly with the number of reference materials.

The third role of ML is to effectively generate new structure candidates. The notable algorithms for this purpose are random search-based algorithms \cite {random_search_3, random_search_1, random_search_2, CALYPSO1, CALYPSO2}, evolutionary-algorithm-based algorithms as USPEX \cite{uspex, uspex_1, uspex_2},  XtalOpt \cite{LONIE2011372}, and recent deep-learning-based models \cite{NOH2019, Ryan18}. In practice, it is possible to generate random structures by forcibly combining different crystal structures $in$ $silico$. The successful discovery of novel material structures under high pressure demonstrates the effectiveness of this approach when certain constraints can be set. However, it is not easy to rationally combine different crystal structures with different compositions and symmetry in a plausible manner. Therefore, oversight in the search for a small number of potential materials cannot be controlled. The combination of first-principles calculations and ML is required for creating effective methods for exploring materials.

One of the most common strategies for generating possible crystal structure candidates is to appropriately combine or apply the atomic substitution method to previously known structures. Beginning with a dataset of host crystal structures with known physical properties and predefined substitution operators, we can employ the atomic substitution method to create new hypothetical crystal structures with the same skeleton as that of the host crystal structures. Widely used substitution operators, such as single-site, multisite, or element substitution operators, are selected depending on the host dataset and experts' suggestions. Typically, these suggestions are based on domain knowledge about the physicochemical similarity between elements, atom--atom interactions, structural stability mechanisms, and target physical property mechanisms. Consequently, the substitution method can work well with knowledge about material synthesis and directly lead to material synthesis ideas. Finally, an $understanding$ of the structure--stability relationship can be directly obtained from screening results, which can help in systematically correcting researchers' suggestions.

\subsection*{Our contribution}
In this study, we propose a protocol for exploring new crystal structures under a given combination of constituent elements and the use of data mining to elucidate the structure--stability relationship (Figure \ref{fig:workflow}). As a demonstration example, we search for the new crystal structures of Nd-Fe-B materials by applying the atomic substitution method to a dataset containing host crystal structures composed of lanthanides, transition metals, and light elements. We apply high-throughput first-principles calculations (Fig. \ref{fig:workflow}, block A) to estimate formation energy. Based on this, we evaluate the phase stability (hereinafter referred to as stability) of all generated Nd-Fe-B crystal structures (section \ref{sec:CH_distance}2.2). The new Nd-Fe-B structures discovered after the screening steps are presented in section \ref{sec:New_structures}2.3. Supervised models are trained to mimic the first-principles calculations from the host and substitution crystal structures and their calculated formation energy. Based on results from supervised learning models, relevance analysis is performed to extract the hidden structural descriptors that determine the formation energy of the generated Nd-Fe-B crystal structures (Fig. \ref{fig:workflow}, block B). Finally, we train an unsupervised learning model  (Fig. \ref{fig:workflow}, block C) that use the obtained relevant descriptors to appropriately group newly generated crystal structures. We compare the obtained group labels and potentially formable states of all crystal structures to determine the relationship between the structure and stability of the Nd-Fe-B crystal structures.

\section{2. Screening for potential formable N\MakeLowercase{d}-F\MakeLowercase{e}-B crystal structures}
\subsection{2.1 Creation of new crystal structure candidates}

In this study, we focus on crystalline magnetic materials comprising lanthanide (LA), transition metal (T), and light (X) atoms. We select the LA atoms from $\{$Y, La, Ce, Pr, Nd, Pm, Sm, Eu, Gd, Tb, Dy, Ho, Er, Tm, Yb, and Lu$\}$; T from $\{$Ti, V, Cr, Mn, Fe, Co, Ni, Cu, Zn, Y, Zr, Nb, Mo, Tc, Ru, Rh, Pd, Ag, Cd, Hf, Ta, W, Re, Os, Ir, Pt, Au, and Hg$\}$; and X from $\{$H, B, C, N, and O$\}$. We collect the details of 5967 well-known crystal structures with formation energies from the Open Quantum Materials Database (OQMD) \cite{OQMD} (version 1.1) to form the host material dataset, which is denoted as $\mathcal{D}_{\text{LA-T-X}}^{host}$.  Each host crystal structure consists of one or two rare-earth metals, one or two transition metals, and one light element. Additionally, from $\mathcal{D}_{\text{LA-T-X}}^{host}$,  we select a subset of all crystal structures comprising Nd, Fe, and B, which is denoted as $\mathcal{D}_{\text{Nd-Fe-B}}^{host}$.

We create new candidates for crystal structures consisting of Nd, Fe, and B with the same skeleton as the host crystal structures in $\mathcal{D}_{\text{LA-T-X}}^{host}$ using a substitution method. For each host crystal structure, a substituted crystal structure is created by substituting all lanthanide sites with Nd, all transition metal sites with Fe, and all light element sites with B. The new structures are compared with each other and with the crystal structures in the $\mathcal{D}_{\text{LA-T-X}}^{host}$ dataset to remove duplication. We follow the comparison procedure proposed by qmpy (the python application programming interface of OQMD) \cite{OQMD}. The structures of the materials are transformed into reduced primitive cells to compare two lattices; all lattice parameters are compared. The internal coordinates of the structures are compared by examining all rotations allowed by each lattice and searching for rotations and translations to map the atoms of the same species into one another within a given level of tolerance. Here, any two structures in which the percent deviation in lattice parameters and angles smaller than 0.1 are considered identical. Further, we apply our designed orbital field matrix (OFM) (section \ref{sec:OFM}3.1) to eliminate duplication. Two structures are considered to be the same if the $L_{2}$ norm of the difference in the OFM is less than $10^{-3}$. Note that two structures that have the same shape but are slightly different in size are considered identical. Finally, we obtain a dataset of the substituted crystal structures, which is denoted as $\mathcal{D}_{\text{Nd-Fe-B}}^{subst}$, with 149 new nonoptimized Nd-Fe-B crystal structures. These structures are then performed structural optimization through first-principles calculations for obtaining the optimal structures.

\begin{figure}[t]
	\includegraphics[width=0.5\textwidth]{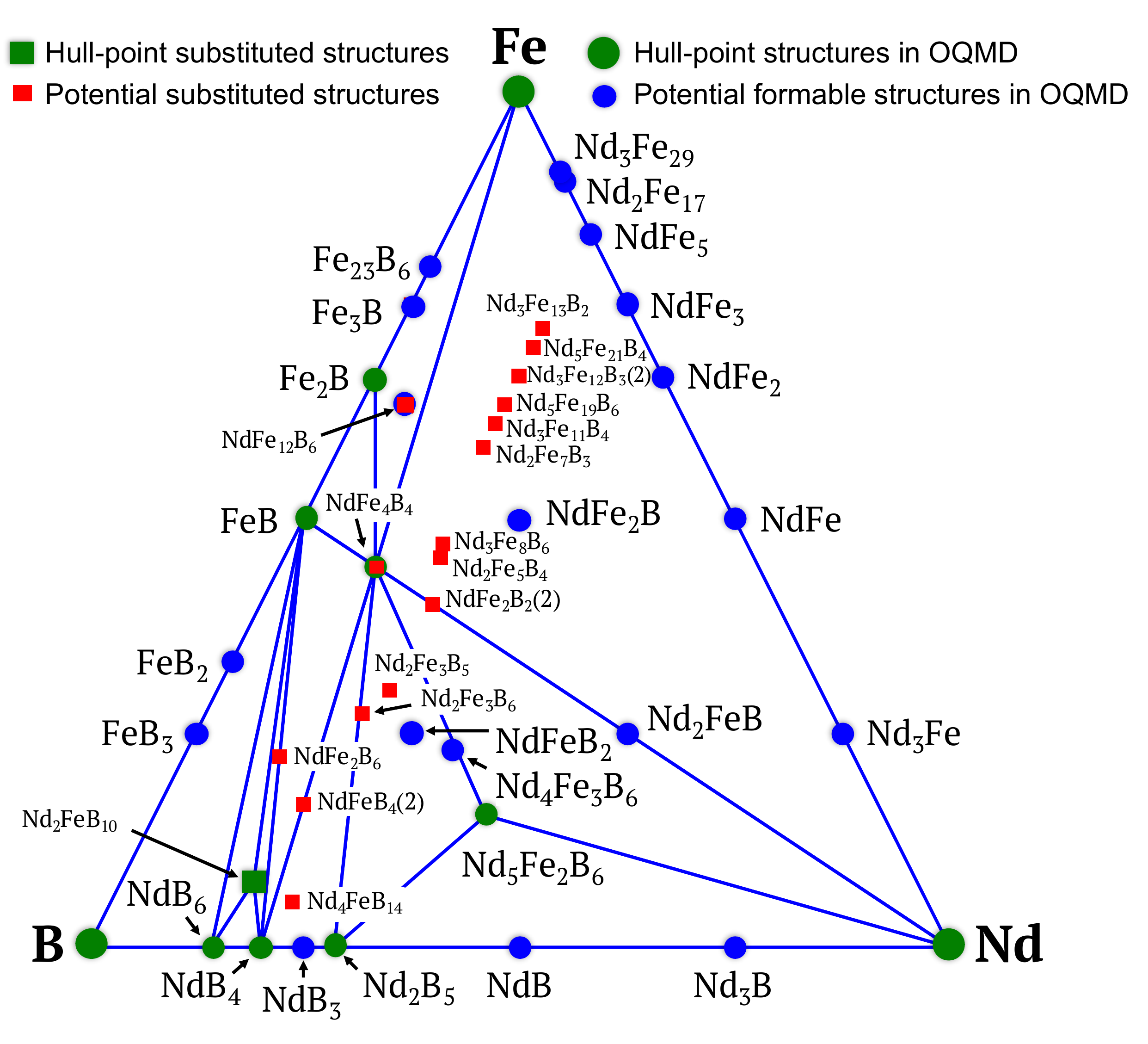}
    \caption{\label{fig:phase_diag} Phase diagram of Nd-Fe-B including materials obtained from OQMD (blue circles) and 20 new substituted structures that confirm it is potentially formable (red squares). Hull points are denoted in green. The total number of disparate structures with the same chemical composition is shown in parentheses.}
\end{figure} 

\begin{table*}[t]
 \caption{\label{tab:1} Properties of new Nd-Fe-B materials: formation energy by DFT $E^{DFT}_f$ (eV/atom), stability by DFT $\Delta E^{DFT}$, magnetization M ($\mu_B$ per formula unit and $\mu_B$ per $\mbox{\AA}^3$ in parentheses), and mean displacement $\Delta r$, estimated by hypothesized structures and final-optimized structures. }
\begin{tabular}{p{22mm}R{18mm}R{18mm}R{22mm}C{18mm}p{22mm}p{22mm}}
\hline\hline
Formula   &\centering  $E_f^{DFT}$ & \centering $\Delta E^{DFT}$& \centering M  &  $\Delta r$ & \multirow{2}{20mm}{Host materials} &  \multirow{2}{20mm}{OQMD id of host materials} \\
          & (eV/atom)  & ( eV/atom) &  ($\mu_B$ ($\mu_B/\mbox{\AA}^3$))  &  ($\mbox{\AA}$) & & \\\hline
Nd$_2$FeB$_{10}$  & -0.522 \, & -0.011 \,  & 13.11 (0.050) & 0.038   & Ce$_2$NiB$_{10}$  & 2025052 \cite{Jeitschko00} \\
NdFe$_2$B$_6$    & -0.473 \, & 0.008 \,        & 3.30  (0.040) & 0.150 &  CeCr$_2$B$_6$  & 94775 \cite{Kuzma72}  \\
Nd$_4$FeB$_{14}$  & -0.506 \, & 0.030 \,   & 26.30 (0.063) & 0.069   & Ho$_4$NiB$_{14}$  &  2107958 \cite{Geupel01}  \\
NdFe$_2$B$_{2}$$-\alpha$    & -0.343 \, & 0.046 \,        & 4.41  (0.067) & 0.085  &  DyCo$_2$B$_2$  &  1852452 \cite{Niihara87}   \\
NdFeB$_{4}$$-\alpha$  & -0.462 \, & 0.052 \,   & 17.42 (0.073) & 0.041   & CeNiB$_4$  & 2023354  \cite{Akselrud84} \\
NdFeB$_{4}$$-\beta$  & -0.455 \, & 0.060 \,   & 18.73 (0.072) & 0.050   & CeCrB$_4$  & 2023373  \cite{Kuzma73-1} \\
Nd$_2$Fe$_3$B$_5$   & -0.374 \, & 0.066 \,     & 6.85  (0.055) & 0.143   & Eu$_2$Os$_3$B$_5$  & 180411 \cite{Jung86}  \\
Nd$_2$Fe$_5$B$_4$   & -0.284 \,  & 0.069 \,    & 10.31 (0.077) & 0.206 & Eu$_2$Rh$_5$B$_4$  & 183842 \cite{Jung90}  \\
NdFe$_4$B$-\alpha$  & -0.092 \, & 0.070 \,     & 21.64 (0.134) & 1.769   & CeCo$_4$B  &  185365 \cite{Kuzma73-2} \\
NdFe$_{12}$B$_{6}$ & -0.231 \, & 0.072 \, & 45.56 (0.117) & 1.012   & CeNi$_{12}$B$_6$  & 2077072  \cite{Akselrud85}\\
Nd$_5$Fe$_{21}$B$_4$  & -0.052 \, & 0.077 \,   & 57.73 (0.140) & 2.342   & Nd$_5$Co$_{21}$B$_4$  & 126928 \cite{Liang01} \\
Nd$_5$Fe$_{19}$B$_6$  & -0.115 \, & 0.080 \,   & 50.02 (0.128) & 1.820   & Nd$_5$Co$_{19}$B$_6$  & 125302 \cite{Liang00} \\
NdFe$_{4}$B$-\beta$  & -0.081 \, & 0.081 \,   & 65.19 (0.135) & 0.241   & NdNi$_4$B  & 2069928 \cite{Noel03} \\
Nd$_3$Fe$_{13}$B$_2$  & -0.027 \, & 0.081 \,   & 36.12 (0.144) & 2.961  & Ce$_3$Ni$_{13}$B$_2$  & 1778822  \cite{Kuzma81}  \\
Nd$_3$Fe$_{11}$B$_4$  & -0.131 \, & 0.085 \,   & 28.22 (0.122) & 0.150   & Ce$_3$Co$_{11}$B$_4$  &  1852403 \cite{Kuzma73-3} \\
Nd$_2$Fe$_3$B$_{6}$  & -0.375 \, & 0.088 \,   & 16.02 (0.066) & 0.132   & Ce$_2$Re$_3$B$_6$  &   1966804 \cite{Kuzma89} \\
NdFe$_{4}$B$_{4}$ & -0.342 \, & 0.090 \, & 17.30 (0.048) & 0.140   & CeRu$_4$B$_4$  & 2074891 \cite{Poettgen10} \\
NdFe$_2$B$_{2}$$-\beta$   & -0.297 \, & 0.092 \,     & 7.25  (0.057) &0.142  & CeIr$_2$B$_2$  & 180315  \cite{Jung91}\\
Nd$_3$Fe$_8$B$_6$   & -0.249 \, & 0.094 \,     & 16.06 (0.079) & 0.543 & Eu$_3$Rh$_8$B$_6$  & 1771853  \cite{Jung90-2} \\
Nd$_2$Fe$_{7}$B$_3$  & -0.147 \, & 0.096 \,   & 35.04 (0.116) & 0.209 & Ce$_2$Co$_7$B$_3$  &  2016489 \cite{Kuzma74} \\
\hline\hline
\end{tabular}
\end{table*}

\subsection{2.2 Assessment of phase stability}\label{sec:CH_distance}
First-principles calculations based on DFT \cite{KS-DFT, HK-DFT} are one of the most effective calculation methods used in materials science. DFT calculations can accurately estimate the formation energy of materials, which is used to build phase diagrams for systems of interest. Hence, the phase stability of a material---in other words, the decomposition energy of a material (CH-distance)---is obtained via the convex-hull analysis of phase diagrams and the decomposition of the material into other phases. We use the formation energy obtained from OQMD \cite{OQMD, OQMD1} of $\mathcal{D}_{\text{LA-T-X}}^{host}$ to build phase diagrams and calculate the CH-distance. The CH-distance of a material is defined as follows:

\begin{equation}
	\label{eq:convex_hull}
	\Delta E = \Delta E_f - E_H,
\end{equation}
where $\Delta E_f$ is the formation energy and $E_H$ is determined by projecting from the chemical composition position to an end point appearing on the convex hull facets. Details of the algorithm for finding these convex hull facets from hull points (see \cite{Barber96,OQMD} for more information). Hereafter, we consider the CH distance $\Delta E$ as the degree of the phase stability of a material. A material that lies below or on the CH surface, $\Delta E = 0$, is a potentially formable material in nature, and a material associated with $\Delta E > 0$ is unstable. A material associated with $\Delta E$ slightly above the CH surface is considered to be in a metastable phase.

Metastable phases are synthesized in numerous cases, for which we consider a reasonable range of the CH distance \cite{AB_perovskite}. Referring to the prediction accuracy of formation energy ($\sim$ 0.1 eV/atom by OQMD \cite{OQMD}), we define all materials with $\Delta E \leq 0.1$ eV/atom as potentially formable structures and as unstable materials otherwise. Following this definition, $\mathcal{D}_{\text{LA-T-X}}^{host}$ can be divided into subsets $\mathcal{D}_{\text{LA-T-X}}^{host\_stb}$ and $\mathcal{D}_{\text{LA-T-X}}^{host\_unstb}$ for potentially formable crystal structures and unstable crystal structures, respectively.

$\mathcal{D}_{\text{Nd-Fe-B}}^{host}$ includes 35 Nd-Fe-B crystal structures, which can be used as references to construct the Nd-Fe-B phase diagram. Seven materials were found for ternary materials, which comprised Nd, Fe, and B. To verify the reliability of the dataset that used to construct phase diagram as well as the stability definition, we remove each tenary materials and use the remaining materials in $\mathcal{D}_{\text{Nd-Fe-B}}^{host}$ to estimate its corresponding convex hull distance. Under this test, among the seven ternary crystal structures, there are two formable ternary materials, NdFe$_4$B$_4$ and Nd$_5$Fe$_2$B$_6$, which lie on the surface of the CH of the phase diagram with $\Delta E = 0.0$. Additionally, one material, NdFe$_{12}$B$_6$, is potentially formable (metastable) with a stability of less than $0.1$ eV/atom, as shown in Table \ref{tab:Nd-Fe-B} in the Supplemental Materials. It should be noted that the important magnetic material, Nd$_2$Fe$_{14}$B, did not exist in the OQMD database at the time when we conducted this study. Based on the Nd-Fe-B phase diagram and the formation energy of -0.057 eV/atom calculated using DFT, the corresponding $\Delta E^{DFT}$ is $1.4 \times 10^{-4}$ eV/atom. This result implies that Nd$_2$Fe$_{14}$B is in the stable phase.  To conclude, we confirm that the experimentally synthesized structures are all satisfy with the stability definition shown in this section. 

We follow the computational settings of OQMD \cite{OQMD, OQMD1} for estimating the formation energy of the newly created Nd-Fe-B crystal structures in $\mathcal{D}_{\text{Nd-Fe-B}}^{subst}$. The calculations were performed using the Vienna ab initio simulation package (VASP) \cite{vasp1,vasp2,vasp3,vasp4} by utilizing projector-augmented wave method potentials (PAW) \cite{paw1, paw2} and the Perdew--Burke--Ernzerhof (PBE)\cite{pbe} exchange-correlation functional.

\begin{figure*}
	\includegraphics[scale=0.4]{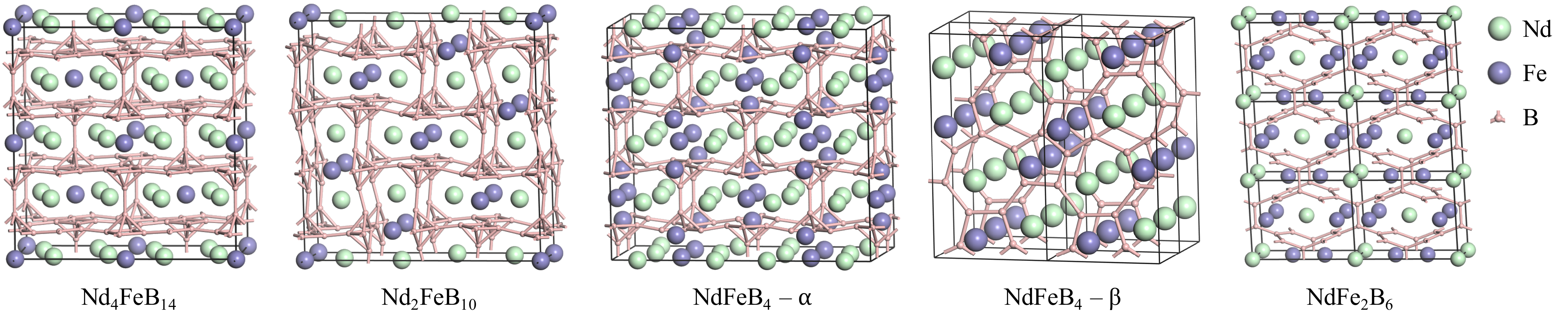}
    \caption{Five representative Nd-Fe-B structures discovered by applying the elemental substitution method to the lanthanides-transition metal and rare-earth material dataset. From left to right: Nd$_4$FeB$_{14}$, Nd$_{2}$FeB$_{10}$, NdFeB$_{4}$$-\alpha$, NdFeB$_{4}$$-\beta$, and NdFe$_2$B$_6$. All twenty discovered structures are shown in the Supplemental Materials.}\label{repr_struct}
\end{figure*}

We employ DFT+U for  Fe, and all calculations are spin-polarized with the ferromagnetic alignment of the spins and with initial magnetic moments of 5, 0, and 0 $\mu_B$ for Fe, Nd, and B, respectively.  For each newly created structure, we perform coarse optimization, fine optimization, and a single point calculation, following the "coarse relax,"  "fine relax," and  "standard" procedures of the OQMD. The k-grid for these calculation series is selected by the k-points per reciprocal atom (KPRA): 4000, 6000, and 8000 for "coarse relax," "fine relax," and  "standard," respectively. We use a cutoff energy of 520 eV for all calculations. The total energies of the "standard" calculations are used for the formation energy calculations, $\Delta E^{DFT}_f$. 
The CH distance of a newly created structure can be estimated from $\Delta E^{DFT} = \Delta E^{DFT}_f - E_H$.

After calculating the formation energy, we finally found 20 new Nd-Fe-B crystal structures that are not in $\mathcal{D}_{\text{LA-T-X}}^{host}$, in which the CH distance of the corresponding optimized structure is less than 0.1 eV. These structures originate from different host structures with different skeletons. Note that we found one structure, Nd$_2$FeB$_{10}$, with a stability of less than $-0.01$ eV/atom. Thus, this structure is also used as a reference to construct the Nd-Fe-B phase diagram. Among the 20 new Nd-Fe-B structures, there are three pairs of discriminated structures sharing the same chemical compositions (NdFe$_2$B$_{2}$, NdFeB$_{4}$ and NdFe$_4$B). Details about these structures are described in Table \ref{tab:1}. The phase diagram of the Nd-Fe-B materials, including the 20 new substituted structures, is shown in Figure \ref{fig:phase_diag}.

We also calculate the magnetization of these materials. We use open-core approximation to treat the 4f electrons of Nd. The contribution of 4f electrons to the magnetization is $J_{g_J} = 3.273$. The magnetization is normalized to the volume of a unit cell:
\begin{equation}
M = M_{DFT} + J_{g_J}\cdot n_{Nd} =  M_{DFT}  + 3.273 \cdot n_{Nd},
\end{equation}
where $M_{DFT}$ is the magnetization given by DFT and $n_{Nd}$ is the number of Nd atoms in the unit cell. All calculation results are summarized in Table \ref{tab:1}. 
 
\subsection{2.3 Newly discovered Nd-Fe-B crystal structures}\label{sec:New_structures}

Figure \ref{repr_struct} shows five specific crystal structures of the predicted formable crystal structures. A common characteristic of these crystal structures is that boron atoms form a network structure and Nd and Fe atoms are surrounded by the cages formed by the boron atom network. In the Nd$_4$FeB$_{14}$ crystal structure, these boron cages are arranged in parallel and Fe atoms are sandwiched between two halves of the boron atom octahedron. In the crystal structure of Nd$_{2}$FeB$_{10}$, which is confirmed by DFT calculations and selected as the hull point in the phase diagram, Nd and Fe atoms are trapped in the boron atom cages; however, these cages are arranged in herringbone patterns. Interestingly, two stable crystal structures of NdFeB$_{4}$ were found as the proportion of Fe increased. One (NdFeB$_{4}$$-\alpha$ structure) was obtained by the elemental substitution of the original CeNiB$_4$ (id 2023354 \cite{Akselrud84}) crystal structure. This crystal structure is similar to the Nd$_4$FeB$_{14}$ crystal structure, with cages formed by boron networks that trap Nd and Fe atoms and are arranged in parallel. In contrast, in the other predicted crystal structure for NdFeB$_4$ (NdFeB$_{4}$$-\beta$ structure obtained by the elemental substitution of the CeCrB$_4$ (id 2023373 \cite{Kuzma73-1}) crystal structure), the boron atoms form a planar network structure that comprises heptagon--pentagon ring pairs. Another form of boron cage is found in NdFe$_2$B$_6$ crystal structure. All potentially formable crystal structures are shown in detail in the Supplemental Materials.

\section{3. Mining structure--stability relationship of N\MakeLowercase{d}-F\MakeLowercase{e}-B crystal structures}\label{sec:ML_models}
\subsection{3.1 Materials representation}\label{sec:OFM}

We must convert the information regarding the materials into descriptor vectors. We employ the OFM \cite{ofm, ofm1} descriptor with a minor modification. The OFM descriptors are constructed using the weighted product of the one-hot vector representations, $\vec{O}$, of atoms. Each vector $\vec{O}$ is filled with zeros, except those representing the electronic configuration of the valence electrons of the corresponding atom. The OFM of a local structure, named $\Theta$, is defined as follows:

\begin{equation}
    \label{eq:ofm_no_d}
     \Theta = \vec{O}^{\top}_{central} \times \left(1.0, \sum_k\frac{\theta_k}{\theta_{max}}\vec{O}_{k}\right),
\end{equation}
where $\theta_k$ is the solid angle determined by the face of the Voronoi polyhedra between the central atom and the index $k$ neighboring atom; and $\theta_{max}$ is the maximum solid angle between the central atom and neighboring atoms. By removing the distance dependence in the original OFM formulation \cite{ofm, ofm1}, we focus exclusively on the coordination of valence orbitals and the shape of the crystal structures. The mean over the local structure descriptors is used as the descriptor of the entire structure:
\begin{equation}
  \label{eq:6} 
\quad OFM_{p} = \frac{1}{N_{p}} \sum_{l=1}^{N_{p}} \Theta_{p}^{l},
\end{equation}
where $p$ is the structure index, and $l$ and $N_p$ are the local structure indices and the number of atoms in the unit cell of the structure $p$, respectively. 

 Note that owing to the designed cross product between the atomic representation vectors of each atom, each element in the matrix represents the average number of atomic coordinates for a certain type of atom. For example, an element of a descriptor obtained by considering the product of the $d^6$ element of the center atom representation and the $f^4$ element of the environment atom representation, denoted as  {$\left(d^6, f^4\right)$}, shows an average coordination number of $f^4$ (Nd) sites surrounding all $d^6$ (Fe) sites. As the term $s^{2}$ appears at all descriptors for Fe, Nd, and B sites, the element {$\left(s^2, s^2\right)$} represent the average coordination number of a given structure. All of these OFM elements provide a foundation for the intuitively interpretable investigation of the structure-stability relationship.

\subsection{3.2 Mining of formation energy data of LA-T-X crystal structures with supervised learning method}\label{sec.mining_ef}

We trained the ML models that can predict the formation energy of the crystal structures, $\Delta E_f$, from $\mathcal{D}_{\text{LA-T-X}}^{host}$, which is represented using the OFM descriptor and the corresponding known formation energy data. We applied kernel ridge regression (KRR)\cite{ML}, which is demonstrated to be useful for predicting material properties. In the  KRR algorithm, the target variable, $y=\Delta E_f$, is represented by a weighted kernel function as follows:
\begin{equation}
\label{eq:krr}
\hat{y}_p = \sum_{k}c_k K(\textbf{x}_p, \textbf{x}_k) = \sum_{k}c_k \exp( {-\gamma |\textbf{x}_p - \textbf{x}_k|)},
\end{equation}
where $\hat{y}_p$ is the predicted formation energy of crystal structure $p$; $\textbf{x}_p$ and $\textbf{x}_k$ are the representation vectors of crystal structures $p$ and $k$ based on the OFM descriptor, respectively; $k$ runs over all crystal structures in the training set; $K(\textbf{x}, \textbf{x}_k)$ is Laplacian kernel function.  Coefficients $c_k$ are estimated by minimizing the total square error regularized by the $L_2$ norm as follows: $\sum_k (\hat{y}_k - y_k)^2 + \lambda \sum_k c_k^2$, with $y_k$ and $\hat{y}_k$ as observed and predicted target values of the structure $k$, respectively.  We perform a ten-times ten-fold cross-validation process to determine parameters $\lambda$ and $\gamma$ in the KRR models. These parameters are selected by minimizing the mean absolute error (MAE) of the validation set.  

\begin{figure}[t]
\centering
  \includegraphics[width=0.8\linewidth]{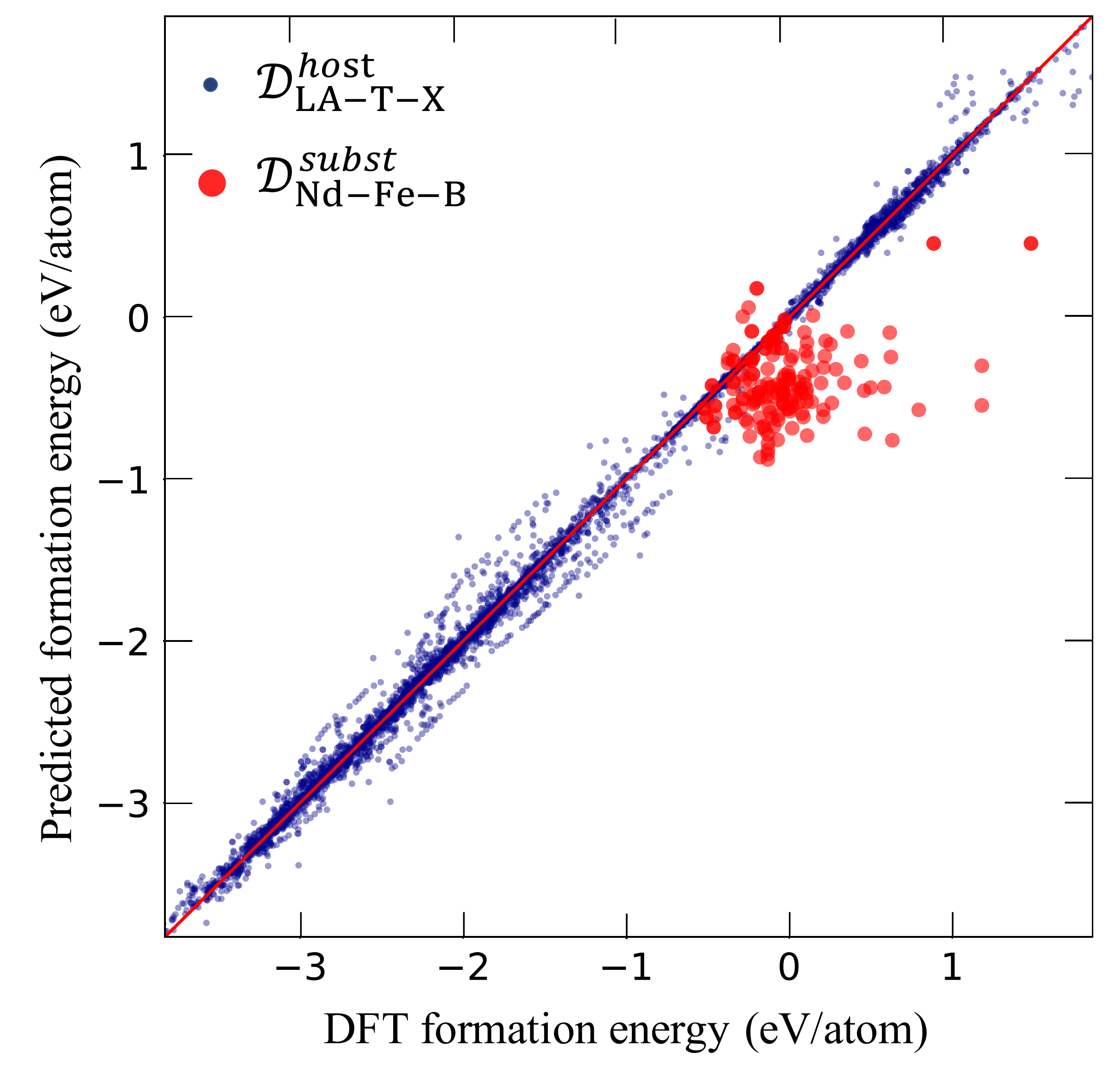}
\caption{Comparison of formation energies calculated using DFT and those predicted through ML  using the KRR model with the OFM descriptor. The blue and red solid circles represent the cross-validated results for $\mathcal{D}_{\text{LA-T-X}}^{host}$ and the prediction results for $\mathcal{D}_{\text{Nd-Fe-B}}^{subst}$, respectively.} \label{fig:krr_results}
\end{figure}

Figure \ref{fig:krr_results} shows the ten-times ten-fold cross-validated comparison of the formation energies calculated using DFT and those predicted by the KRR model for the crystal structures in $\mathcal{D}_{\text{LA-T-X}}^{host}$ (blue circles). Figure \ref{fig:krr_results} also shows a comparison of the formation energies calculated using DFT and those predicted using the KRR model (trained using all crystal structures in $\mathcal{D}_{\text{LA-T-X}}^{host}$) for the crystal structures in   $\mathcal{D}_{\text{Nd-Fe-B}}^{subst}$ (red circles). In the cross-validated comparison of materials in $\mathcal{D}_{\text{LA-T-X}}^{host}$, the  formation energies predicted via KRR show good agreement with those calculated using DFT, with $R^2$ \cite{Kvalseth85} value of  0.990(1), table \ref{tab:ML_cv}. 

\begin{table}[t]
\caption{\label{tab:ML_cv} Ten-times ten-fold cross-validation results provided by the KRR model in predicting formation energy.}
\begin{tabular}{C{20mm}C{18mm}C{18mm}C{18mm} } 
\hline\hline
Model 	& 	$R^2$  & MAE (eV/atom) 	&  RMSE (eV/atom) \\   \hline \hline                            
Kernel ridge 	&  0.990(1)   &  0.094(2)   &  0.137(1)	\\ \hline
\hline
\end{tabular}
\end{table}

It should be noted that this predictive model is learned from the data ($\mathcal{D}_{\text{LA-T-X}}^{host}$) containing only the optimized crystal structures. Thus, when applied to a newly generated nonoptimized crystal structure (in $\mathcal{D}_{\text{Nd-Fe-B}}^{subst}$), it is clear that the possibility of correctly predicting the  formation energy is low. The MAE of the KRR-predicted formation energy of the crystal structures in $\mathcal{D}_{\text{Nd-Fe-B}}^{subst}$ after structure optimization is approximately 0.3 (eV/atom), which is three times larger than the cross-validated MAE result. The results of applying the KRR prediction model to estimate the stability of these hypothetical materials are shown in detail in section \ref{sec:Ther}3.5.

\begin{figure}[t]
\centering
  \includegraphics[width=0.95\linewidth]{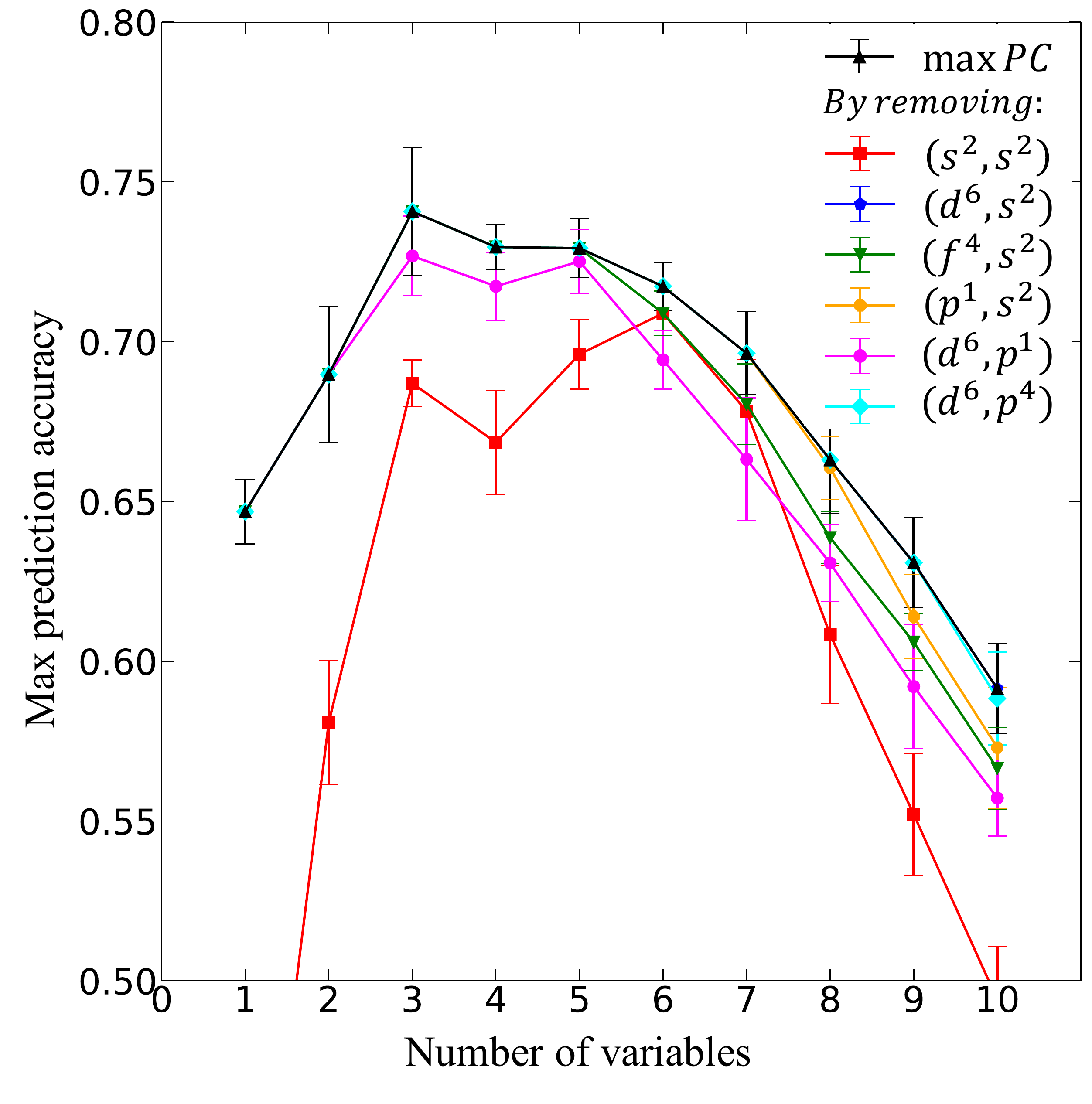}
\caption{Results of relevance analysis performed for predicting $E_{f}$ of all Nd-Fe-B materials present in $\mathcal{D}_{\text{LA-T-X}}^{host}$. By removing the descriptor $(s^2, s^2)$, the maximum predictioncapacity (red line) significantly reduces compared to the maximum prediction capacity line (max $PC$) of all descriptor sets (black line). }\label{fig:PA_result}
\end{figure}

\subsection{3.3 Descriptor relevant analysis}\label{sec.method_relevant}

Further, we focus on $\mathcal{D}_{\text{Nd-Fe-B}}^{host}$ and evaluate the relevance \cite{Nguyen_2019, feature_selection_Yu:2004, feature_selection} of each element in the OFM descriptor with respect to the formation energy of the crystal structure. We utilize the change in prediction accuracy when removing or adding a descriptor (from the full set of descriptors \cite{Nguyen18} in the OFM) to search for the descriptors that are strongly relevant \cite{Nguyen_2019, Dam18} to the formation energy (i.e., CH distance and phase stability) of the Nd-Fe-B crystal structures.

In detail, for a given set $S$ of descriptors, we define the prediction capacity $PC(S)$ of $S$ by the maximum prediction accuracy that the KRR model can achieve by using the variables in a subset $s$ of $S$ as follows:
\begin{equation}
PC(S) = \max_{\forall s \subset S} R^2_s; \\  s_{PC} = \argmax_{\forall s \subset S} R^2_s, 
\label{eq.PAS}
\end{equation}
where $R^2_s$ is the value of the coefficient of determination $R^2$ \cite{Kvalseth85} achieved by the KRR using a set $s$ as the independent variables. $s_{PA}$ is the subset of $S$ that yields the prediction model having the maximum prediction accuracy.

Let denote $S_i$ as set of descriptors after removing a descriptor $x_i$ from the full descriptor set $S$; $S_i = S - \{x_i$\}.  A descriptor is strongly relevant if and only if:
\begin{equation}
PC(S)-PC(S_i) = \max_{\forall s \subset S} R^2_s - \max_{\forall s \subset S_i} R^2_s > 0.
\label{eq.strong}
\end{equation}
\begin{figure}[t]
\centering
\includegraphics[width=1.05\linewidth]{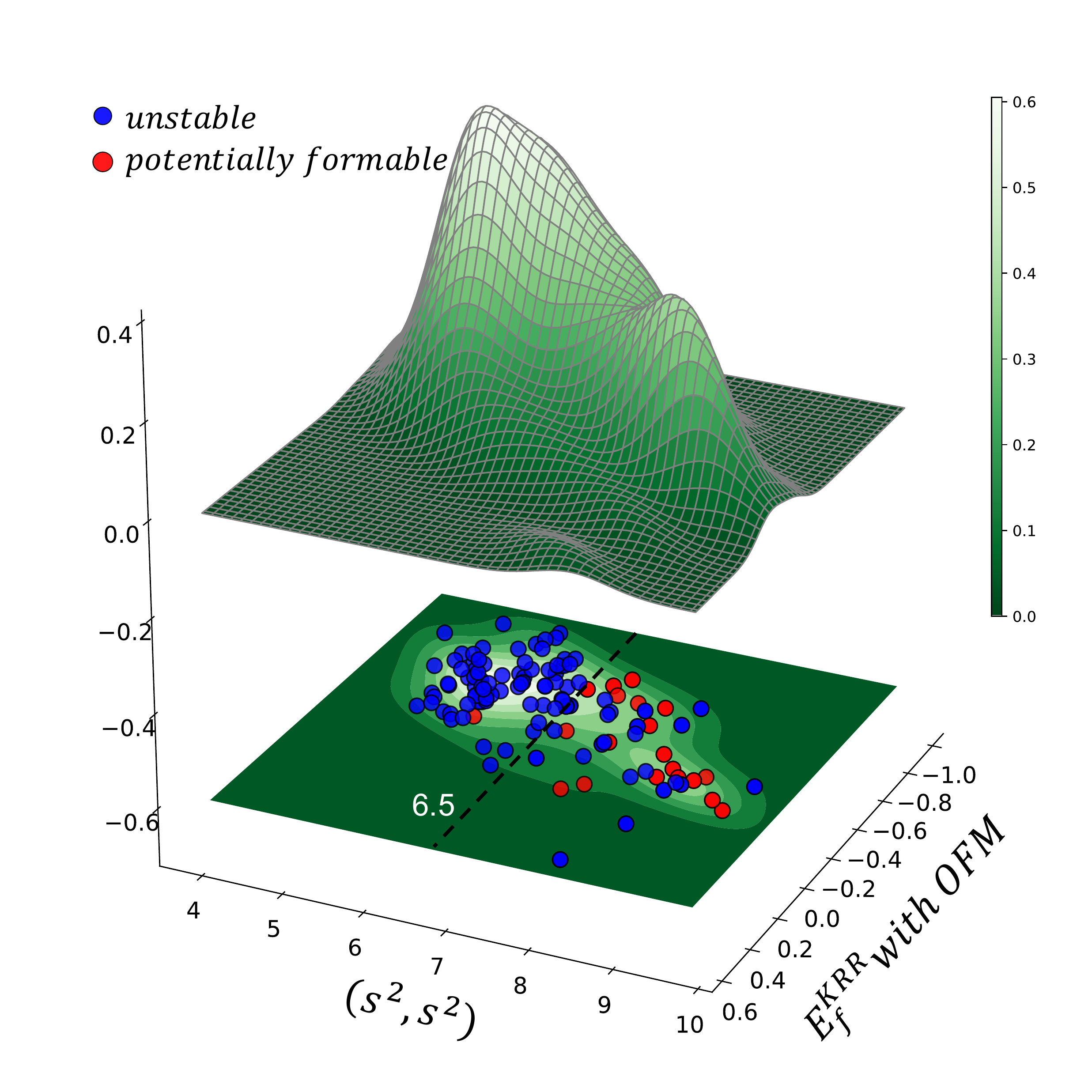}
\caption{Distribution of substituted materials in the space with $x$-axis showing the KRR-predicted formation energy, $E_f^{KRR}$, with nonoptimized structures and $y$-axis showing the extracted strongly relevant descriptor $(s^2, s^2)$. The black dotted line shows the limitation of $(s^2, s^2)$, which maximizes the separation between two mixture distributions.}\label{fig:clf_result}
\end{figure}
Figure \ref{fig:PA_result} summarizes the results obtained from the descriptor relevance analysis. The black--triangled curve shows the dependence of the max prediction capacity (max $PC$, in $R^2$ score) on the number of variables--OFM descriptors that used in regression models. Other curves show the dependence of the max prediction capacity on the number of OFM  that used in regression models when specific OFM is removed from whole set of OFM descriptors. For example, the orange--dot curve illustrates the max $PC$ of the OFM descriptor set without appearance of $(p^1, s^2)$ descriptor. It is evident that the descriptor $(s^2, s^2)$ (red--square curve) is highly relevant to the prediction of the formation energy of the crystal structures in $\mathcal{D}_{\text{Nd-Fe-B}}^{subst}$. For further investigation, we project all substituted crystal structures in $\mathcal{D}_{\text{Nd-Fe-B}}^{subst}$ into the space of the KRR-predicted formation energy, $E_{f}^{KRR}$, and $(s^2, s^2)$, as shown in Figure \ref{fig:clf_result}. One can easily deduce that the distribution of $\mathcal{D}_{\text{Nd-Fe-B}}^{subst}$ is a mixture of two distribution components. The larger distribution component is located in the region $(s^2, s^2) < 6.5$, whereas the other is located in the region $(s^2, s^2) \geq 6.5$. We infer the existence of two distinct groups of substituted crystal structures. The first group contains structures with average atomic coordination numbers lower than 6.5, and the second group contains structures with average atomic coordination numbers higher than 6.5. Further, most newly discovered potentially formable crystal structures belong to the second group. 

\begin{figure*}
	\includegraphics[scale=0.56]{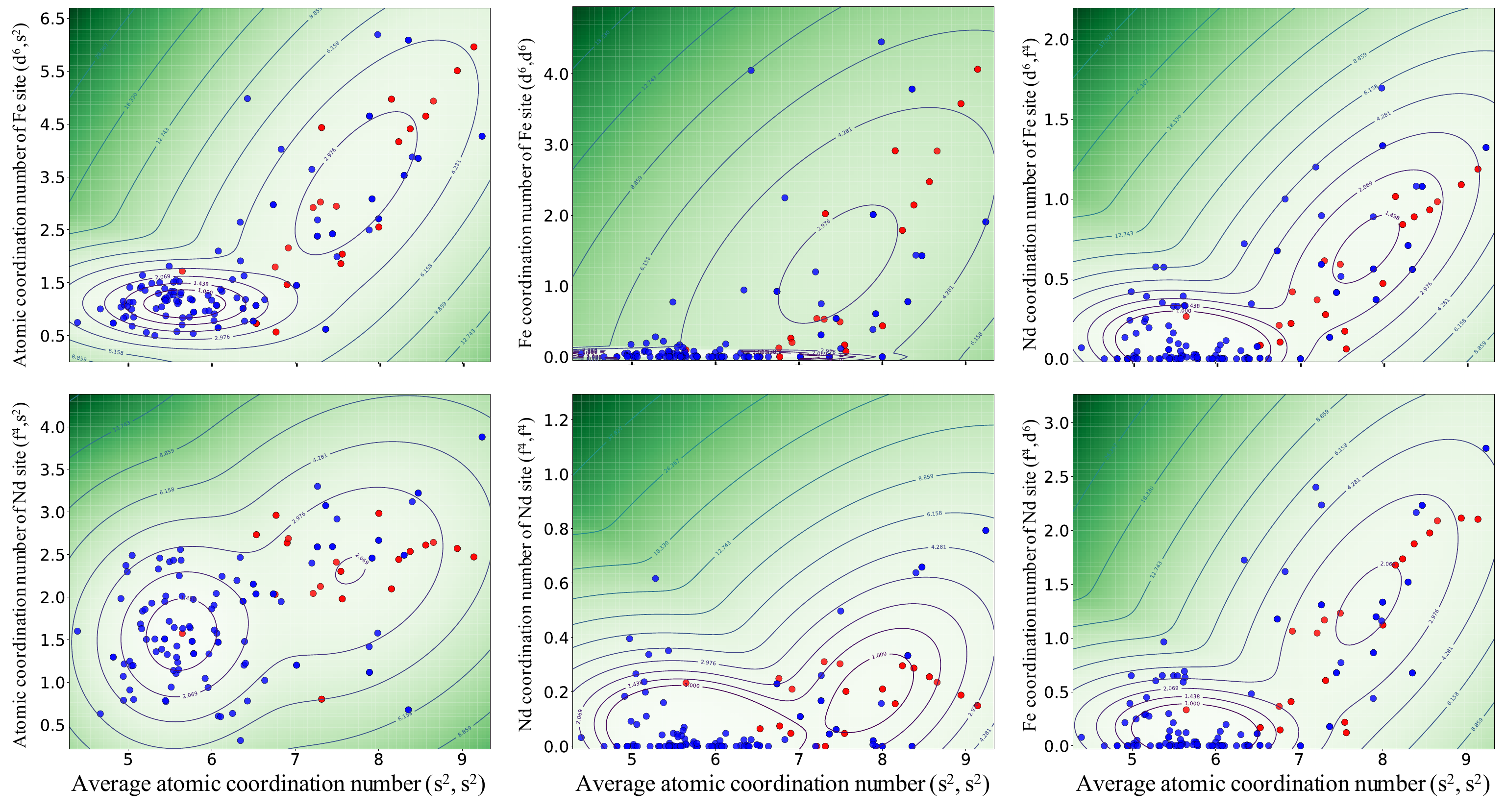}
    \caption{\label{fig:Distribution}Density distribution of newly generated Nd-Fe-B crystal structures in two-dimensional space obtained using selected OFM descriptors. The blue and red solid circles represent the unstable and potentially formable crystal structures verified by DFT calculations, respectively. Contour lines show the isodense  surface of the distribution.}
\end{figure*}

\subsection{3.4 Mining of substituted Nd-Fe-B crystal structure data with unsupervised learning method}

In this section, we demonstrate  the use of the proposed generative model, which applies the relevance analysis results and unsupervised learning, in contrast to the conventional supervised learning approach. As a result, this model performs detailed investigations at particular sites whose coordination numbers are highly correlated to the structure--stability relationship.

The underlying hypothesis of this approach is that there are various correlation patterns between crystal structure properties and their formation energies. Naturally, most of these patterns are for unstable crystal structures and only a few of these pertain to potentially formable crystal structures. These patterns might not be exposed directly through feature relevance analysis method due to the multivariate correlation between the target and predicting variables. The strong relevant descriptor $(s^2, s^2)$ can appear as an extracted pattern to indicate the correlation between the structure--stability relationship. As the term $s^{2}$ appears at all descriptors for Fe, Nd, and B sites, $(s^2, s^2)$  indicates only the average atomic coordination numbers, which do not precisely represent the coordination number of any particular site. On the contrary, other OFM descriptors are designed to explicitly represent the coordination number of all pairwise elements. As the two terms $d^{6}$ and $f^{4}$ appears at only descriptors for Fe or Nd, respectively. Therefore, to investigate the average coordination number of the Fe, Nd, and B sites, in addition to $(s^2, s^2)$, we focus on the values of the descriptors $(d^6, s^2)$, $(f^4, s^2)$. These descriptors represent the average atomic coordination numbers of Fe sites and Nd sites. Further, we also focus on the values of the OFM descriptors $(d^6, d^6)$, $(d^6, f^4)$, $(f^4, d^6)$, and $(f^4, f^4)$. These descriptors represent the average number of Fe sites surrounding the Fe sites, Nd sites surrounding the Fe sites, Fe sites surrounding the Nd sites, and Nd sites surrounding the Nd sites. These OFM descriptors are useful in discussing not only the structure--stability relationship but also the strength of magnetic exchange couplings  between the 3$d$ orbitals of Fe and the 4$f$ orbitals of Nd.

Figure \ref{fig:Distribution} shows the density distribution of the newly created crystal structures, $\mathcal{D}_{\text{Nd-Fe-B}}^{subst}$, in two-dimensional space using the selected descriptors. For all pairs of descriptor, the density distribution is similar to the distribution of $(s^2, s^2)$ and $E_{f}^{KRR}$ shown in Figure \ref{fig:clf_result} with two clear peaks, one large and one small, with slight overlap. This result again confirms that $(s^2, s^2)$ is highly relevant for expressing the nature of distribution of the newly created crystal structures. In addition, $(d^6, s^2)$ and $(d^6, d^6)$ are important for identifying the characteristics of the distribution. It should be noted that, these features could not be exposed by using feature relevance analysis since the prediction model can utilize the information from other highly correlated feature instead e.g. $(s^s, s^2)$. In contrast, the average coordination number of the Nd sites $(f^4, s^2)$ and the average coordination number of the Nd sites around the Nd sites $(f^4, f^4)$ have a weak relationship with the characteristics related to the distribution of these crystal structures. These results indicate that the average coordination number of the Fe sites $(d^6, s^2)$ and the average coordination number of the Fe sites around the Fe sites $(d^6, d^6)$ are extremely important for characterizing the newly created Nd-Fe-B crystal structures.

We employ a GMM \cite{ML} for learning the patterns of crystal structures by clustering $\mathcal{D}_{\text{Nd-Fe-B}}^{subst}$ into groups. The GMM model is based on the assumptions that the data consist of different groups and the data in each group follow their own Gaussian distribution. In other words, in the GMM, the distribution of data is fitted to a combination of a certain number, $M$, of Gaussian functions~\cite{ML} with $M$ represents for the number of data groups. The probability distribution of a crystal structure with index $p$, represented using selected descriptors, $\textbf{x}_{p}$ and $f(\textbf{x}_{p})$, can be approximated as follows:

\begin{equation}\label{eq.1}
f(\textbf{x}_{p}) = \sum_{m=1}^{M}\alpha_m \Phi({\textbf{x}}_{p}, \bm{\mu}_m, \bm{\Sigma}_m),
\end{equation}
where 
\begin{equation}\label{eq. 2}
\Phi(\textbf{x}_{p}, \bm{\mu}_m, \bm{\Sigma}_m) = 
\frac{\exp \left[-(\textbf{x}_{p}-\bm{\mu}_m)^T \bm{\Sigma}_m^{-1}(\textbf{x}_{p} - \bm{\mu}_m)\right]}{(2\pi)^{\frac{d}{2}}|\bm{\Sigma}_m|^{\frac{1}{2}}}
\end{equation}

is a multivariate Gaussian distribution with mean $\bm{\mu}_m$ and covariance matrix $\bm{\Sigma}_m$ and $d$ is the dimension of representation vector $\textbf{x}_{p}$. Coefficients $\alpha_m$ are the weights that satisfy the following constraint:

\begin{equation}\label{eq. 3}
\sum_{m=1}^M\alpha_m = 1. 
\end{equation}

The probability that $\textbf{x}_{p}$ belongs to group $m$ can be represented as follows: 

\begin{equation}\label{eq.4}
p(\textbf{x}_{p} | m) = \frac{\alpha_m \Phi(\textbf{x}_{p}, \bm{\mu}_m, \bm{\Sigma}_m)}{\sum_{m=1}^{M}\alpha_m \Phi(\textbf{x}_{p}, \bm{\mu}_m, \bm{\Sigma}_m)}.
\end{equation}

The model parameters, $\alpha_m, \bm{\mu}_m, \bm{\Sigma}_m$, are determined using an expectation-maximization algorithm \cite{scikit-learn}. The number of data groups, $M$, is fixed at two in this study.  It is interesting to note that the GMM provides a "probabilistic image" of the pattern of the crystal structures, wherein it provides the probability of a crystal structure remaining in a group instead of assigning the crystal structures to a specific group. The sum of the probabilities of crystal structures remaining in either of the groups is 1. Therefore, the GMM is expected to discover distinctive patterns of crystal structures from the data and calculate the probability that a crystal structure belongs to a group. 

We can label the newly generated crystal structures by fitting the data $\mathcal{D}_{\text{Nd-Fe-B}}^{subst}$ to the GMM with two Gaussian distributions and calculating the probabilities of the crystal structures belonging to each group. Given that it is not easy to find a new potential formable crystal structure, we suppose that most newly generated structures are unstable and only a few are potential formable. Therefore, we infer that the large Gauss component corresponds to the distribution of unstable crystal structures and the small Gauss component corresponds to the distribution of potential formable crystal structures. This hypothesis can be verified through comparison with the results of the DFT calculations, and it can be seen that most of the potential formable crystal structures confirmed by the DFT calculation actually belong to the small Gauss component. This implies that the phase stabilities of the Nd-Fe-B crystals are not significantly related to the coordination number of the Nd sites but are largely determined by the coordination number of the Fe sites. This suggests that if the Nd sites can be replaced partly by Fe, the crystal structure characteristics of Nd-Fe-B that are directly related to its  phase stability can be controlled. Further application of this discovery in the design of Nd-Fe-B crystal materials is promising.

\begin{table}[t]
\caption{Evaluation results of KRR, LG, DT models, and unsupervised GMM in estimating the stability of materials in $\mathcal{D}_{\text{Nd-Fe-B}}^{subst}$.}\label{tab:results_accuracy}
\begin{tabular}{p{25mm}p{18mm}p{16mm}p{13mm} }
\hline\hline
Model     & \hfil $Precision$ & \hfil $Recall$ & \hfil $f_{1}$   \\ \hline
KRR model & \hfil 0.533     & \hfil 0.534  & \hfil 0.376 \\ \hline
LG-model  & \hfil 0.629     & \hfil 0.687  & \hfil 0.599 \\ \hline
DT-model  & \hfil 0.704     & \hfil 0.676  & \hfil 0.687 \\ \hline
GMM & \hfil \textbf{0.729}     & \hfil \textbf{0.821}  & \hfil \textbf{0.735} \\ \hline\hline
\end{tabular}
\end{table}

 \begin{table*}[t]
 \caption{Classification results in predicting "potentially formable" class label of substituted materials with KRR, LG, DT models, GMM, and ensemble models. The AND and OR operators in these ensemble models are denoted  by $"\&"$ and $"|"$, respectively}.
 \label{tab:GMM_pred_results}
 \begin{tabular}{|c|c|c|c|c|c|c|c|c|c|c|c}
 \hline \hline 
 		& KRR  & LG  & DT  & GMM  & KRR$|$GMM  & LG$|$GMM  & DT$|$GMM  & KRR $\&$ GMM & LG $\&$ GMM & DT $\&$ GMM \\ \hline
  Precision & 0.24  & 0.35  & 0.56  & 0.49  & 0.24  & 0.36  & 0.48 & \textbf{0.58} & 0.53 & \textbf{0.58} \\ \hline
  Recall 		& 0.82 & 0.79 & 0.45 & 0.91 & 0.97  & \textbf{1.0} & 0.91 & 0.76 & 0.7 & 0.45  \\ \hline
  $f_1$ & 0.37 & 0.49 & 0.5 & 0.64 & 0.39  & 0.53 & 0.63  & \textbf{0.66} & 0.61 &0.51  \\ \hline
\hline 
\end{tabular}

 \caption{Classification results in predicting "unstable" class label of substituted materials with KRR, LG, DT models, GMM, and ensemble models. The AND and OR operators in these ensemble models are denoted  by $"\&"$ and $"|"$, respectively.}\label{tab:GMM_pred_results_2}

\begin{tabular}{|c|c|c|c|c|c|c|c|c|c|c|c|}
 \hline \hline 
 		& KRR  & LG  & DT  & GMM  & KRR$|$GMM  & LG$|$GMM  & DT$|$GMM  & KRR $\&$ GMM & LG $\&$ GMM & DT $\&$ GMM \\ \hline
  Precision & 0.83  & 0.91 & 0.85 & 0.97 & 0.94 & \textbf{1.0} & 0.97 & 0.92 & 0.91 & 0.85 \\ \hline
  Recall 		& 0.25 & 0.59 & 0.90 & 0.73 & 0.14  & 0.49 & 0.72 & 0.84 & 0.83 & \textbf{0.91} \\ \hline
  $f_1$ & 0.38 & 0.71 & 0.87 & 0.83 & 0.24 & 0.66 & 0.83  & \textbf{0.88} & 0.86 & 0\textbf{.88} \\ \hline
\hline 
\end{tabular}

\end{table*}

\subsection{3.5 Learning prediction models for phase stability of crystal structures} \label{sec:Ther}
A large number of ML applications reported until now (in materials science research) state the effectiveness and applicability of ML methods using statistical tests (such as cross validation). However, statistical tests are methods for assessing the risk in predicting the physical properties of the most optimized structure materials, and are they not appropriate for predicting and discovering novel materials. Therefore, in this study, to verify whether ML techniques are effective in searching for new potentially formable Nd-Fe-B crystal structures, we train three supervised ML models from $\mathcal{D}_{\text{LA-T-X}}^{host}$ and one unsupervised model from $\mathcal{D}_{\text{Nd-Fe-B}}^{subst}$. Additionally, we test if the models can predict the stability of the newly created crystal structures in $\mathcal{D}_{\text{Nd-Fe-B}}^{subst}$. The three supervised ML models are trained by considering 5967 materials in $\mathcal{D}_{\text{LA-T-X}}^{host}$ with the OFM descriptor and the stability target values described in sections 3.1 and 2.2. Then, all models are applied to predict the 149  newly hypothetical structures in $\mathcal{D}_{\text{Nd-Fe-B}}^{subst}$ while considering the stability calculated by the DFT as references in prediction accuracy evaluation.

In the first model (KRR model), the CH-distance is calculated using the formation energy predicted by the KRR model described in section 3.2. Then, we apply a threshold of 0.1 eV/atom to the obtained CH-distance to determine whether the crystal structure is potentially formable. It is worth noting again that the bottleneck of this method is that the formation energy prediction model is learned from data containing only the optimal crystal structures. Therefore, formation energy is not predicted correctly when the method is applied to a newly created nonoptimal crystal structure.

The second model is a logistic regression model (LG-model). From the two subsets of $\mathcal{D}_{\text{LA-T-X}}^{host}$, including the potentially formable ($\mathcal{D}_{\text{LA-T-X}}^{host\_stb}$) and unstable ($\mathcal{D}_{\text{LA-T-X}}^{host\_unstb}$) crystal structures, we model the probability of observing potentially formable ($y = 1$) and unstable ($y = 0$) class labels directly using classification models. We hypothesize the probability of observing potentially formable materials, $p(y=1 | X = \textbf{x}_p)$, as follows:
\begin{equation}
\label{eq:LG}
    p(y=1 | X = \textbf{x}_p) = \frac{\exp \sum_{i} c_i x_{pi}}  {1 + \exp \sum_{i} c_i x_{pi} }, 
\end{equation} 
where $\textbf{x}_{p}$ is the description vector of structure $p$ (obtained by flattening the OFM), $i$ is the index of vector elements in $\textbf{x}_{p}$, and $c_i$ is the coefficient of the corresponding element, $x_{pi}$. In our experiments, all coefficients, $c_i$, are obtained via maximum a posteriori estimation using $L_1$ as the regularization term \cite{ANg04, Lee06}. The third model is the decision tree model (DT-model) \cite{ML}, which uses information gain \cite{Breiman84, hastie2009elements}  as the criterion to measure the quality of tree splitting.

The unsupervised model is based on the observations of the mixture distribution of the newly created crystal structures, $\mathcal{D}_{\text{Nd-Fe-B}}^{subst}$. We build the fourth model (GMM) by assuming that the obtained major and minor Gauss components  correspond to the "unstable" and "potentially formable" class labels of the crystal structures, respectively.

The evaluation results of the four models are summarized in Table \ref{tab:results_accuracy}. We use  three evaluation scores: $Precision$, $Recall$, and $f_1$. The $Precision$ score (also referred to as positive predictive value) with respect to the unstable structure class is the fraction of the unstable crystal structures that are predicted correctly among the number of crystal structures that are predicted to be unstable\cite{Perry55}. The $Recall$ score (also known as sensitivity) with respect to the unstable structure class is the fraction of the unstable crystal structures that are predicted correctly among all crystal structures that are actually unstable\cite{Perry55}. The $Precision$ and $Recall$ scores are combined in the $f_1$ score (or f-measure) to provide a single measurement\cite{Derczynski16}. To compare the classification ability of ML models, we summarize the evaluation scores of all classes (i.e., "unstable" and "potentially formable") by utilizing a macro averaging method\cite{Su15}, which is implemented in $\rm{sklearn.metrics.average\_precision\_score}$\cite{scikit-learn} version 0.21.3. 


The KRR model shows the lowest values of all evaluation scores among the three supervised learning models with   $Precision$, $Recall$ and $f_1$ are 0.533, 0.534 and 0.376, respectively.  In contrast, the DT-model provides the most accurate prediction. This model accurately predicts the potentially formable-unstable label of all substituted Nd-Fe-B crystal structures with 0.704 macro $Precision$ score and obtains macro $Recall$ and $f_1$ scores of 0.676 and 0.687, respectively. The LG-model shows the highest macro $Recall$ score, 0.687, compared with the other two supervised learning models. 

The final but most surprising result is that the unsupervised GMM is superior to the other three supervised learning models in all three evaluation scores. The average $Precision$ and $Recall$ scores of the GMM are 0.729 and 0.821, respectively, which are significantly higher than those of the three supervised learning models. This result shows that the integration of descriptor relevance analysis and unsupervised learning with the GMM is superior to conventional ML models, such as KRR, LG, and DT, for obtaining information about the  phase stability of the substituted Nd-Fe-B crystal structures. We also investigate the usefulness of ensembling models. As the prediction problem under consideration is a binary classification, we implement two well-known operators, "AND" and "OR," for combining classification results. The details of the results are shown in Tables \ref{tab:GMM_pred_results} and \ref{tab:GMM_pred_results_2}. These results again suggest that the structure--stability relationship obtained using data mining is highly promising for the design of Nd-Fe-B materials.

\section{4. Conclusion}

We focus on discovering new Nd-Fe-B materials using the elemental substitution method with LA-T-X compounds---lanthanide, transition metal, and light element (X = B, C, N, O)---as host materials. For each host crystal structure, a substituted crystal structure is created by substituting all lanthanide sites with Nd, all transition metal sites with Fe, and all light element sites with B. High-throughput first-principles calculations are applied to evaluate the phase stability of the newly created crystal structures, and twenty of them are found to be potential formable. We implemented an approach by incorporating supervised and unsupervised learning techniques to estimate the stability and analyze the relationship between the structure and stability of the newly created Nd-Fe-B crystal structures. Three supervised learning models (KRR, LG, and DT models) learned from LA-T-X host crystal structures achieve the maximum accuracy and recall scores of $70.4\%$ and $68.7\%$, respectively, in predicting the stability state of the new substituted Nd-Fe-B crystals. The proposed unsupervised learning model resulting from the integration of descriptor-relevance analysis and the GMM provides accuracy and recall scores of $72.9\%$ and $82.1\%$, respectively, which are significantly better than those of the supervised models. Moreover, the unsupervised learning model can capture and interpret the structure--stability relationship of the Nd-Fe-B crystal structures. The average atomic coordination number and the coordination number of the Fe sites are quantitatively shown to be the most important factors in determining the phase stability of the new substituted Nd-Fe-B crystal structures.

\section*{acknowledgments}
\begin{acknowledgments}

This work is supported by the Precursory Research for Embryonic Science and Technology from the Japan Science and Technology Agency (JST), ESICMM Grant Number 12016013 (ESICMM is funded by Ministry of Education, Culture, Sports, Science and Technology (MEXT)), JSPS KAKENHI Grants Number 20K05301 and Number JP19H05815 (Grant-in-Aid for Scientific Research on Innovative Areas “Interface Ionics”), Materials Research by Information Integration Initiative (MI$^2$I) project of the Support Program for Starting Up Innovation Hub from JST, and MEXT as a social and scientific priority issue employing the post-K computer (creation of new functional devices and high-performance materials to support next-generation industries; CDMSI).

\end{acknowledgments}

\bibliography{main}

\begin{thebibliography}{83}%
\makeatletter
\providecommand \@ifxundefined [1]{%
 \@ifx{#1\undefined}
}%
\providecommand \@ifnum [1]{%
 \ifnum #1\expandafter \@firstoftwo
 \else \expandafter \@secondoftwo
 \fi
}%
\providecommand \@ifx [1]{%
 \ifx #1\expandafter \@firstoftwo
 \else \expandafter \@secondoftwo
 \fi
}%
\providecommand \natexlab [1]{#1}%
\providecommand \enquote  [1]{``#1''}%
\providecommand \bibnamefont  [1]{#1}%
\providecommand \bibfnamefont [1]{#1}%
\providecommand \citenamefont [1]{#1}%
\providecommand \href@noop [0]{\@secondoftwo}%
\providecommand \href [0]{\begingroup \@sanitize@url \@href}%
\providecommand \@href[1]{\@@startlink{#1}\@@href}%
\providecommand \@@href[1]{\endgroup#1\@@endlink}%
\providecommand \@sanitize@url [0]{\catcode `\\12\catcode `\$12\catcode
  `\&12\catcode `\#12\catcode `\^12\catcode `\_12\catcode `\%12\relax}%
\providecommand \@@startlink[1]{}%
\providecommand \@@endlink[0]{}%
\providecommand \url  [0]{\begingroup\@sanitize@url \@url }%
\providecommand \@url [1]{\endgroup\@href {#1}{\urlprefix }}%
\providecommand \urlprefix  [0]{URL }%
\providecommand \Eprint [0]{\href }%
\providecommand \doibase [0]{http://dx.doi.org/}%
\providecommand \selectlanguage [0]{\@gobble}%
\providecommand \bibinfo  [0]{\@secondoftwo}%
\providecommand \bibfield  [0]{\@secondoftwo}%
\providecommand \translation [1]{[#1]}%
\providecommand \BibitemOpen [0]{}%
\providecommand \bibitemStop [0]{}%
\providecommand \bibitemNoStop [0]{.\EOS\space}%
\providecommand \EOS [0]{\spacefactor3000\relax}%
\providecommand \BibitemShut  [1]{\csname bibitem#1\endcsname}%
\let\auto@bib@innerbib\@empty
\bibitem [{\citenamefont {Butler}\ \emph {et~al.}(2018)\citenamefont {Butler},
  \citenamefont {Davies}, \citenamefont {Cartwright}, \citenamefont {Isayev},\
  and\ \citenamefont {Walsh}}]{Butler18}%
  \BibitemOpen
  \bibfield  {author} {\bibinfo {author} {\bibfnamefont {K.~T.}\ \bibnamefont
  {Butler}}, \bibinfo {author} {\bibfnamefont {D.~W.}\ \bibnamefont {Davies}},
  \bibinfo {author} {\bibfnamefont {H.}~\bibnamefont {Cartwright}}, \bibinfo
  {author} {\bibfnamefont {O.}~\bibnamefont {Isayev}}, \ and\ \bibinfo {author}
  {\bibfnamefont {A.}~\bibnamefont {Walsh}},\ }\href@noop {} {\bibfield
  {journal} {\bibinfo  {journal} {Nature}\ }\textbf {\bibinfo {volume} {559}},\
  \bibinfo {pages} {547} (\bibinfo {year} {2018})}\BibitemShut {NoStop}%
\bibitem [{\citenamefont {Curtarolo}\ \emph {et~al.}(2013)\citenamefont
  {Curtarolo}, \citenamefont {Hart}, \citenamefont {Nardelli}, \citenamefont
  {Mingo}, \citenamefont {Sanvito},\ and\ \citenamefont {Levy}}]{Curtarolo13}%
  \BibitemOpen
  \bibfield  {author} {\bibinfo {author} {\bibfnamefont {S.}~\bibnamefont
  {Curtarolo}}, \bibinfo {author} {\bibfnamefont {G.~L.~W.}\ \bibnamefont
  {Hart}}, \bibinfo {author} {\bibfnamefont {M.~B.}\ \bibnamefont {Nardelli}},
  \bibinfo {author} {\bibfnamefont {N.}~\bibnamefont {Mingo}}, \bibinfo
  {author} {\bibfnamefont {S.}~\bibnamefont {Sanvito}}, \ and\ \bibinfo
  {author} {\bibfnamefont {O.}~\bibnamefont {Levy}},\ }\href@noop {} {\bibfield
   {journal} {\bibinfo  {journal} {Nature Materials}\ }\textbf {\bibinfo
  {volume} {12}},\ \bibinfo {pages} {191 EP } (\bibinfo {year}
  {2013})}\BibitemShut {NoStop}%
\bibitem [{\citenamefont {Saal}\ \emph
  {et~al.}(2013{\natexlab{a}})\citenamefont {Saal}, \citenamefont {Kirklin},
  \citenamefont {Aykol}, \citenamefont {Meredig},\ and\ \citenamefont
  {Wolverton}}]{Saal2013}%
  \BibitemOpen
  \bibfield  {author} {\bibinfo {author} {\bibfnamefont {J.~E.}\ \bibnamefont
  {Saal}}, \bibinfo {author} {\bibfnamefont {S.}~\bibnamefont {Kirklin}},
  \bibinfo {author} {\bibfnamefont {M.}~\bibnamefont {Aykol}}, \bibinfo
  {author} {\bibfnamefont {B.}~\bibnamefont {Meredig}}, \ and\ \bibinfo
  {author} {\bibfnamefont {C.}~\bibnamefont {Wolverton}},\ }\href {\doibase
  10.1007/s11837-013-0755-4} {\bibfield  {journal} {\bibinfo  {journal} {JOM}\
  }\textbf {\bibinfo {volume} {65}},\ \bibinfo {pages} {1501} (\bibinfo {year}
  {2013}{\natexlab{a}})}\BibitemShut {NoStop}%
\bibitem [{\citenamefont {Korner}\ \emph {et~al.}(2016)\citenamefont {Korner},
  \citenamefont {Krugel},\ and\ \citenamefont {Elsasser}}]{Korner16}%
  \BibitemOpen
  \bibfield  {author} {\bibinfo {author} {\bibfnamefont {W.}~\bibnamefont
  {Korner}}, \bibinfo {author} {\bibfnamefont {G.}~\bibnamefont {Krugel}}, \
  and\ \bibinfo {author} {\bibfnamefont {C.}~\bibnamefont {Elsasser}},\ }\href
  {\doibase doi:10.1038/srep24686} {\bibfield  {journal} {\bibinfo  {journal}
  {Scientific Reports}\ }\textbf {\bibinfo {volume} {6}} (\bibinfo {year}
  {2016}),\ doi:10.1038/srep24686}\BibitemShut {NoStop}%
\bibitem [{\citenamefont {Korner}\ \emph {et~al.}(2018)\citenamefont {Korner},
  \citenamefont {Krugel}, \citenamefont {Urban},\ and\ \citenamefont
  {Elsasser}}]{KORNER18}%
  \BibitemOpen
  \bibfield  {author} {\bibinfo {author} {\bibfnamefont {W.}~\bibnamefont
  {Korner}}, \bibinfo {author} {\bibfnamefont {G.}~\bibnamefont {Krugel}},
  \bibinfo {author} {\bibfnamefont {D.~F.}\ \bibnamefont {Urban}}, \ and\
  \bibinfo {author} {\bibfnamefont {C.}~\bibnamefont {Elsasser}},\ }\href
  {\doibase https://doi.org/10.1016/j.scriptamat.2017.11.038} {\bibfield
  {journal} {\bibinfo  {journal} {Scripta Materialia}\ }\textbf {\bibinfo
  {volume} {154}},\ \bibinfo {pages} {295 } (\bibinfo {year}
  {2018})}\BibitemShut {NoStop}%
\bibitem [{\citenamefont {Ma}\ \emph {et~al.}(2017)\citenamefont {Ma},
  \citenamefont {Hegde}, \citenamefont {Munira}, \citenamefont {Xie},
  \citenamefont {Keshavarz}, \citenamefont {Mildebrath}, \citenamefont
  {Wolverton}, \citenamefont {Ghosh},\ and\ \citenamefont
  {Butler}}]{PhysRevB.95.024411}%
  \BibitemOpen
  \bibfield  {author} {\bibinfo {author} {\bibfnamefont {J.}~\bibnamefont
  {Ma}}, \bibinfo {author} {\bibfnamefont {V.~I.}\ \bibnamefont {Hegde}},
  \bibinfo {author} {\bibfnamefont {K.}~\bibnamefont {Munira}}, \bibinfo
  {author} {\bibfnamefont {Y.}~\bibnamefont {Xie}}, \bibinfo {author}
  {\bibfnamefont {S.}~\bibnamefont {Keshavarz}}, \bibinfo {author}
  {\bibfnamefont {D.~T.}\ \bibnamefont {Mildebrath}}, \bibinfo {author}
  {\bibfnamefont {C.}~\bibnamefont {Wolverton}}, \bibinfo {author}
  {\bibfnamefont {A.~W.}\ \bibnamefont {Ghosh}}, \ and\ \bibinfo {author}
  {\bibfnamefont {W.~H.}\ \bibnamefont {Butler}},\ }\href {\doibase
  10.1103/PhysRevB.95.024411} {\bibfield  {journal} {\bibinfo  {journal} {Phys.
  Rev. B}\ }\textbf {\bibinfo {volume} {95}},\ \bibinfo {pages} {024411}
  (\bibinfo {year} {2017})}\BibitemShut {NoStop}%
\bibitem [{\citenamefont {He}\ \emph {et~al.}(2018)\citenamefont {He},
  \citenamefont {Naghavi}, \citenamefont {Hegde}, \citenamefont {Amsler},\ and\
  \citenamefont {Wolverton}}]{Jiangang18}%
  \BibitemOpen
  \bibfield  {author} {\bibinfo {author} {\bibfnamefont {J.}~\bibnamefont
  {He}}, \bibinfo {author} {\bibfnamefont {S.~S.}\ \bibnamefont {Naghavi}},
  \bibinfo {author} {\bibfnamefont {V.~I.}\ \bibnamefont {Hegde}}, \bibinfo
  {author} {\bibfnamefont {M.}~\bibnamefont {Amsler}}, \ and\ \bibinfo {author}
  {\bibfnamefont {C.}~\bibnamefont {Wolverton}},\ }\href {\doibase
  10.1021/acs.chemmater.8b01096} {\bibfield  {journal} {\bibinfo  {journal}
  {Chemistry of Materials}\ }\textbf {\bibinfo {volume} {30}},\ \bibinfo
  {pages} {4978} (\bibinfo {year} {2018})},\ \Eprint
  {http://arxiv.org/abs/https://doi.org/10.1021/acs.chemmater.8b01096}
  {https://doi.org/10.1021/acs.chemmater.8b01096} \BibitemShut {NoStop}%
\bibitem [{\citenamefont {Balluff}\ \emph {et~al.}(2017)\citenamefont
  {Balluff}, \citenamefont {Diekmann}, \citenamefont {Reiss},\ and\
  \citenamefont {Meinert}}]{Balluff17}%
  \BibitemOpen
  \bibfield  {author} {\bibinfo {author} {\bibfnamefont {J.}~\bibnamefont
  {Balluff}}, \bibinfo {author} {\bibfnamefont {K.}~\bibnamefont {Diekmann}},
  \bibinfo {author} {\bibfnamefont {G.}~\bibnamefont {Reiss}}, \ and\ \bibinfo
  {author} {\bibfnamefont {M.}~\bibnamefont {Meinert}},\ }\href {\doibase
  10.1103/PhysRevMaterials.1.034404} {\bibfield  {journal} {\bibinfo  {journal}
  {Phys. Rev. Materials}\ }\textbf {\bibinfo {volume} {1}},\ \bibinfo {pages}
  {034404} (\bibinfo {year} {2017})}\BibitemShut {NoStop}%
\bibitem [{\citenamefont {Yang}\ \emph {et~al.}(2012)\citenamefont {Yang},
  \citenamefont {Setyawan}, \citenamefont {Wang}, \citenamefont
  {Buongiorno~Nardelli},\ and\ \citenamefont {Curtarolo}}]{Yang11}%
  \BibitemOpen
  \bibfield  {author} {\bibinfo {author} {\bibfnamefont {K.}~\bibnamefont
  {Yang}}, \bibinfo {author} {\bibfnamefont {W.}~\bibnamefont {Setyawan}},
  \bibinfo {author} {\bibfnamefont {S.}~\bibnamefont {Wang}}, \bibinfo {author}
  {\bibfnamefont {M.}~\bibnamefont {Buongiorno~Nardelli}}, \ and\ \bibinfo
  {author} {\bibfnamefont {S.}~\bibnamefont {Curtarolo}},\ }\href@noop {}
  {\bibfield  {journal} {\bibinfo  {journal} {Nature Materials}\ }\textbf
  {\bibinfo {volume} {11}},\ \bibinfo {pages} {614} (\bibinfo {year}
  {2012})}\BibitemShut {NoStop}%
\bibitem [{\citenamefont {Li}\ \emph {et~al.}(2018)\citenamefont {Li},
  \citenamefont {Zhang}, \citenamefont {Yao},\ and\ \citenamefont
  {Zhang}}]{Li_2018}%
  \BibitemOpen
  \bibfield  {author} {\bibinfo {author} {\bibfnamefont {X.}~\bibnamefont
  {Li}}, \bibinfo {author} {\bibfnamefont {Z.}~\bibnamefont {Zhang}}, \bibinfo
  {author} {\bibfnamefont {Y.}~\bibnamefont {Yao}}, \ and\ \bibinfo {author}
  {\bibfnamefont {H.}~\bibnamefont {Zhang}},\ }\href {\doibase
  10.1088/2053-1583/aadb1e} {\bibfield  {journal} {\bibinfo  {journal} {2D
  Materials}\ }\textbf {\bibinfo {volume} {5}},\ \bibinfo {pages} {045023}
  (\bibinfo {year} {2018})}\BibitemShut {NoStop}%
\bibitem [{\citenamefont {Emery}\ \emph {et~al.}(2016)\citenamefont {Emery},
  \citenamefont {Saal}, \citenamefont {Kirklin}, \citenamefont {Hegde},\ and\
  \citenamefont {Wolverton}}]{Emery16}%
  \BibitemOpen
  \bibfield  {author} {\bibinfo {author} {\bibfnamefont {A.~A.}\ \bibnamefont
  {Emery}}, \bibinfo {author} {\bibfnamefont {J.~E.}\ \bibnamefont {Saal}},
  \bibinfo {author} {\bibfnamefont {S.}~\bibnamefont {Kirklin}}, \bibinfo
  {author} {\bibfnamefont {V.~I.}\ \bibnamefont {Hegde}}, \ and\ \bibinfo
  {author} {\bibfnamefont {C.}~\bibnamefont {Wolverton}},\ }\href {\doibase
  10.1021/acs.chemmater.6b01182} {\bibfield  {journal} {\bibinfo  {journal}
  {Chemistry of Materials}\ }\textbf {\bibinfo {volume} {28}},\ \bibinfo
  {pages} {5621} (\bibinfo {year} {2016})},\ \Eprint
  {http://arxiv.org/abs/https://doi.org/10.1021/acs.chemmater.6b01182}
  {https://doi.org/10.1021/acs.chemmater.6b01182} \BibitemShut {NoStop}%
\bibitem [{\citenamefont {Michalsky}\ and\ \citenamefont
  {Steinfeld}(2017)}]{MICHALSKY2017124}%
  \BibitemOpen
  \bibfield  {author} {\bibinfo {author} {\bibfnamefont {R.}~\bibnamefont
  {Michalsky}}\ and\ \bibinfo {author} {\bibfnamefont {A.}~\bibnamefont
  {Steinfeld}},\ }\href {\doibase https://doi.org/10.1016/j.cattod.2016.09.023}
  {\bibfield  {journal} {\bibinfo  {journal} {Catalysis Today}\ }\textbf
  {\bibinfo {volume} {286}},\ \bibinfo {pages} {124 } (\bibinfo {year}
  {2017})},\ \bibinfo {note} {nitrogen Activation}\BibitemShut {NoStop}%
\bibitem [{\citenamefont {Aykol}\ \emph {et~al.}(2016)\citenamefont {Aykol},
  \citenamefont {Kim}, \citenamefont {Hegde}, \citenamefont {Snydacker},
  \citenamefont {Lu}, \citenamefont {Hao}, \citenamefont {Kirklin},
  \citenamefont {Morgan},\ and\ \citenamefont {Wolverton}}]{Aykol16}%
  \BibitemOpen
  \bibfield  {author} {\bibinfo {author} {\bibfnamefont {M.}~\bibnamefont
  {Aykol}}, \bibinfo {author} {\bibfnamefont {S.}~\bibnamefont {Kim}}, \bibinfo
  {author} {\bibfnamefont {V.~I.}\ \bibnamefont {Hegde}}, \bibinfo {author}
  {\bibfnamefont {D.}~\bibnamefont {Snydacker}}, \bibinfo {author}
  {\bibfnamefont {Z.}~\bibnamefont {Lu}}, \bibinfo {author} {\bibfnamefont
  {S.}~\bibnamefont {Hao}}, \bibinfo {author} {\bibfnamefont {S.}~\bibnamefont
  {Kirklin}}, \bibinfo {author} {\bibfnamefont {D.}~\bibnamefont {Morgan}}, \
  and\ \bibinfo {author} {\bibfnamefont {C.}~\bibnamefont {Wolverton}},\
  }\href@noop {} {\bibfield  {journal} {\bibinfo  {journal} {Nature
  Communications}\ }\textbf {\bibinfo {volume} {7}},\ \bibinfo {pages} {13779}
  (\bibinfo {year} {2016})}\BibitemShut {NoStop}%
\bibitem [{\citenamefont {Ashton}\ \emph {et~al.}(2016)\citenamefont {Ashton},
  \citenamefont {Hennig}, \citenamefont {Broderick}, \citenamefont {Rajan},\
  and\ \citenamefont {Sinnott}}]{Ashton16}%
  \BibitemOpen
  \bibfield  {author} {\bibinfo {author} {\bibfnamefont {M.}~\bibnamefont
  {Ashton}}, \bibinfo {author} {\bibfnamefont {R.~G.}\ \bibnamefont {Hennig}},
  \bibinfo {author} {\bibfnamefont {S.~R.}\ \bibnamefont {Broderick}}, \bibinfo
  {author} {\bibfnamefont {K.}~\bibnamefont {Rajan}}, \ and\ \bibinfo {author}
  {\bibfnamefont {S.~B.}\ \bibnamefont {Sinnott}},\ }\href {\doibase
  10.1103/PhysRevB.94.054116} {\bibfield  {journal} {\bibinfo  {journal} {Phys.
  Rev. B}\ }\textbf {\bibinfo {volume} {94}},\ \bibinfo {pages} {054116}
  (\bibinfo {year} {2016})}\BibitemShut {NoStop}%
\bibitem [{\citenamefont {Moller}\ \emph {et~al.}(2018)\citenamefont {Moller},
  \citenamefont {Korner}, \citenamefont {Krugel}, \citenamefont {Urban},\ and\
  \citenamefont {Elsasser}}]{MOLLER18}%
  \BibitemOpen
  \bibfield  {author} {\bibinfo {author} {\bibfnamefont {J.~J.}\ \bibnamefont
  {Moller}}, \bibinfo {author} {\bibfnamefont {W.}~\bibnamefont {Korner}},
  \bibinfo {author} {\bibfnamefont {G.}~\bibnamefont {Krugel}}, \bibinfo
  {author} {\bibfnamefont {D.~F.}\ \bibnamefont {Urban}}, \ and\ \bibinfo
  {author} {\bibfnamefont {C.}~\bibnamefont {Elsasser}},\ }\href {\doibase
  https://doi.org/10.1016/j.actamat.2018.03.051} {\bibfield  {journal}
  {\bibinfo  {journal} {Acta Materialia}\ }\textbf {\bibinfo {volume} {153}},\
  \bibinfo {pages} {53 } (\bibinfo {year} {2018})}\BibitemShut {NoStop}%
\bibitem [{\citenamefont {Kim}\ \emph {et~al.}(2018)\citenamefont {Kim},
  \citenamefont {Ward}, \citenamefont {He}, \citenamefont {Krishna},
  \citenamefont {Agrawal},\ and\ \citenamefont
  {Wolverton}}]{PhysRevMaterials.2.123801}%
  \BibitemOpen
  \bibfield  {author} {\bibinfo {author} {\bibfnamefont {K.}~\bibnamefont
  {Kim}}, \bibinfo {author} {\bibfnamefont {L.}~\bibnamefont {Ward}}, \bibinfo
  {author} {\bibfnamefont {J.}~\bibnamefont {He}}, \bibinfo {author}
  {\bibfnamefont {A.}~\bibnamefont {Krishna}}, \bibinfo {author} {\bibfnamefont
  {A.}~\bibnamefont {Agrawal}}, \ and\ \bibinfo {author} {\bibfnamefont
  {C.}~\bibnamefont {Wolverton}},\ }\href {\doibase
  10.1103/PhysRevMaterials.2.123801} {\bibfield  {journal} {\bibinfo  {journal}
  {Phys. Rev. Materials}\ }\textbf {\bibinfo {volume} {2}},\ \bibinfo {pages}
  {123801} (\bibinfo {year} {2018})}\BibitemShut {NoStop}%
\bibitem [{\citenamefont {Ulissi}\ \emph {et~al.}(2017)\citenamefont {Ulissi},
  \citenamefont {Tang}, \citenamefont {Xiao}, \citenamefont {Liu},
  \citenamefont {Torelli}, \citenamefont {Karamad}, \citenamefont {Cummins},
  \citenamefont {Hahn}, \citenamefont {Lewis}, \citenamefont {Jaramillo},
  \citenamefont {Chan},\ and\ \citenamefont {Nørskov}}]{Ulissi17}%
  \BibitemOpen
  \bibfield  {author} {\bibinfo {author} {\bibfnamefont {Z.~W.}\ \bibnamefont
  {Ulissi}}, \bibinfo {author} {\bibfnamefont {M.~T.}\ \bibnamefont {Tang}},
  \bibinfo {author} {\bibfnamefont {J.}~\bibnamefont {Xiao}}, \bibinfo {author}
  {\bibfnamefont {X.}~\bibnamefont {Liu}}, \bibinfo {author} {\bibfnamefont
  {D.~A.}\ \bibnamefont {Torelli}}, \bibinfo {author} {\bibfnamefont
  {M.}~\bibnamefont {Karamad}}, \bibinfo {author} {\bibfnamefont
  {K.}~\bibnamefont {Cummins}}, \bibinfo {author} {\bibfnamefont
  {C.}~\bibnamefont {Hahn}}, \bibinfo {author} {\bibfnamefont {N.~S.}\
  \bibnamefont {Lewis}}, \bibinfo {author} {\bibfnamefont {T.~F.}\ \bibnamefont
  {Jaramillo}}, \bibinfo {author} {\bibfnamefont {K.}~\bibnamefont {Chan}}, \
  and\ \bibinfo {author} {\bibfnamefont {J.~K.}\ \bibnamefont {Nørskov}},\
  }\href {\doibase 10.1021/acscatal.7b01648} {\bibfield  {journal} {\bibinfo
  {journal} {ACS Catalysis}\ }\textbf {\bibinfo {volume} {7}},\ \bibinfo
  {pages} {6600} (\bibinfo {year} {2017})}\BibitemShut {NoStop}%
\bibitem [{\citenamefont {Xue}\ \emph {et~al.}(2016{\natexlab{a}})\citenamefont
  {Xue}, \citenamefont {Balachandran}, \citenamefont {Yuan}, \citenamefont
  {Hu}, \citenamefont {Qian}, \citenamefont {Dougherty},\ and\ \citenamefont
  {Lookman}}]{Xue13301}%
  \BibitemOpen
  \bibfield  {author} {\bibinfo {author} {\bibfnamefont {D.}~\bibnamefont
  {Xue}}, \bibinfo {author} {\bibfnamefont {P.~V.}\ \bibnamefont
  {Balachandran}}, \bibinfo {author} {\bibfnamefont {R.}~\bibnamefont {Yuan}},
  \bibinfo {author} {\bibfnamefont {T.}~\bibnamefont {Hu}}, \bibinfo {author}
  {\bibfnamefont {X.}~\bibnamefont {Qian}}, \bibinfo {author} {\bibfnamefont
  {E.~R.}\ \bibnamefont {Dougherty}}, \ and\ \bibinfo {author} {\bibfnamefont
  {T.}~\bibnamefont {Lookman}},\ }\href {\doibase 10.1073/pnas.1607412113}
  {\bibfield  {journal} {\bibinfo  {journal} {Proceedings of the National
  Academy of Sciences}\ }\textbf {\bibinfo {volume} {113}},\ \bibinfo {pages}
  {13301} (\bibinfo {year} {2016}{\natexlab{a}})},\ \Eprint
  {http://arxiv.org/abs/https://www.pnas.org/content/113/47/13301.full.pdf}
  {https://www.pnas.org/content/113/47/13301.full.pdf} \BibitemShut {NoStop}%
\bibitem [{\citenamefont {Mannodi-Kanakkithodi}\ \emph
  {et~al.}(2016)\citenamefont {Mannodi-Kanakkithodi}, \citenamefont {Pilania},
  \citenamefont {Huan}, \citenamefont {Lookman},\ and\ \citenamefont
  {Ramprasad}}]{Mannodi16}%
  \BibitemOpen
  \bibfield  {author} {\bibinfo {author} {\bibfnamefont {A.}~\bibnamefont
  {Mannodi-Kanakkithodi}}, \bibinfo {author} {\bibfnamefont {G.}~\bibnamefont
  {Pilania}}, \bibinfo {author} {\bibfnamefont {T.~D.}\ \bibnamefont {Huan}},
  \bibinfo {author} {\bibfnamefont {T.}~\bibnamefont {Lookman}}, \ and\
  \bibinfo {author} {\bibfnamefont {R.}~\bibnamefont {Ramprasad}},\ }\href@noop
  {} {\bibfield  {journal} {\bibinfo  {journal} {Scientific Reports}\ ,\
  \bibinfo {pages} {20952 EP}} (\bibinfo {year} {2016})}\BibitemShut {NoStop}%
\bibitem [{\citenamefont {Pilania}\ \emph {et~al.}(2016)\citenamefont
  {Pilania}, \citenamefont {Balachandran}, \citenamefont {Kim},\ and\
  \citenamefont {Lookman}}]{Pilania16}%
  \BibitemOpen
  \bibfield  {author} {\bibinfo {author} {\bibfnamefont {G.}~\bibnamefont
  {Pilania}}, \bibinfo {author} {\bibfnamefont {P.~V.}\ \bibnamefont
  {Balachandran}}, \bibinfo {author} {\bibfnamefont {C.}~\bibnamefont {Kim}}, \
  and\ \bibinfo {author} {\bibfnamefont {T.}~\bibnamefont {Lookman}},\ }\href
  {\doibase 10.3389/fmats.2016.00019} {\bibfield  {journal} {\bibinfo
  {journal} {Frontiers in Materials}\ }\textbf {\bibinfo {volume} {3}},\
  \bibinfo {pages} {19} (\bibinfo {year} {2016})}\BibitemShut {NoStop}%
\bibitem [{\citenamefont {Xue}\ \emph {et~al.}(2016{\natexlab{b}})\citenamefont
  {Xue}, \citenamefont {Balachandran}, \citenamefont {Hogden}, \citenamefont
  {Theiler}, \citenamefont {Xue},\ and\ \citenamefont {Lookman}}]{Xue16}%
  \BibitemOpen
  \bibfield  {author} {\bibinfo {author} {\bibfnamefont {D.}~\bibnamefont
  {Xue}}, \bibinfo {author} {\bibfnamefont {P.~V.}\ \bibnamefont
  {Balachandran}}, \bibinfo {author} {\bibfnamefont {J.}~\bibnamefont
  {Hogden}}, \bibinfo {author} {\bibfnamefont {J.}~\bibnamefont {Theiler}},
  \bibinfo {author} {\bibfnamefont {D.}~\bibnamefont {Xue}}, \ and\ \bibinfo
  {author} {\bibfnamefont {T.}~\bibnamefont {Lookman}},\ }\href@noop {}
  {\bibfield  {journal} {\bibinfo  {journal} {Nature Communications}\ }\textbf
  {\bibinfo {volume} {7}},\ \bibinfo {pages} {11241} (\bibinfo {year}
  {2016}{\natexlab{b}})}\BibitemShut {NoStop}%
\bibitem [{\citenamefont {Yamashita}\ \emph {et~al.}(2018)\citenamefont
  {Yamashita}, \citenamefont {Sato}, \citenamefont {Kino}, \citenamefont
  {Miyake}, \citenamefont {Tsuda},\ and\ \citenamefont {Oguchi}}]{Yamashita18}%
  \BibitemOpen
  \bibfield  {author} {\bibinfo {author} {\bibfnamefont {T.}~\bibnamefont
  {Yamashita}}, \bibinfo {author} {\bibfnamefont {N.}~\bibnamefont {Sato}},
  \bibinfo {author} {\bibfnamefont {H.}~\bibnamefont {Kino}}, \bibinfo {author}
  {\bibfnamefont {T.}~\bibnamefont {Miyake}}, \bibinfo {author} {\bibfnamefont
  {K.}~\bibnamefont {Tsuda}}, \ and\ \bibinfo {author} {\bibfnamefont
  {T.}~\bibnamefont {Oguchi}},\ }\href {\doibase
  10.1103/PhysRevMaterials.2.013803} {\bibfield  {journal} {\bibinfo  {journal}
  {Phys. Rev. Materials}\ }\textbf {\bibinfo {volume} {2}},\ \bibinfo {pages}
  {013803} (\bibinfo {year} {2018})}\BibitemShut {NoStop}%
\bibitem [{\citenamefont {Pickard}\ and\ \citenamefont
  {Needs}(2011)}]{random_search_3}%
  \BibitemOpen
  \bibfield  {author} {\bibinfo {author} {\bibfnamefont {C.~J.}\ \bibnamefont
  {Pickard}}\ and\ \bibinfo {author} {\bibfnamefont {R.~J.}\ \bibnamefont
  {Needs}},\ }\href@noop {} {\bibfield  {journal} {\bibinfo  {journal} {Journal
  of Physics: Condensed Matter}\ }\textbf {\bibinfo {volume} {23}},\ \bibinfo
  {pages} {053201} (\bibinfo {year} {2011})}\BibitemShut {NoStop}%
\bibitem [{\citenamefont {Pickard}\ and\ \citenamefont
  {Needs}(2006)}]{random_search_1}%
  \BibitemOpen
  \bibfield  {author} {\bibinfo {author} {\bibfnamefont {C.~J.}\ \bibnamefont
  {Pickard}}\ and\ \bibinfo {author} {\bibfnamefont {R.~J.}\ \bibnamefont
  {Needs}},\ }\href@noop {} {\bibfield  {journal} {\bibinfo  {journal} {Phys.
  Rev. Lett.}\ }\textbf {\bibinfo {volume} {97}},\ \bibinfo {pages} {045504}
  (\bibinfo {year} {2006})}\BibitemShut {NoStop}%
\bibitem [{\citenamefont {Pickard}\ and\ \citenamefont
  {Needs}(2007)}]{random_search_2}%
  \BibitemOpen
  \bibfield  {author} {\bibinfo {author} {\bibfnamefont {C.~J.}\ \bibnamefont
  {Pickard}}\ and\ \bibinfo {author} {\bibfnamefont {R.~J.}\ \bibnamefont
  {Needs}},\ }\href@noop {} {\bibfield  {journal} {\bibinfo  {journal} {Nature
  Physics}\ }\textbf {\bibinfo {volume} {3}} (\bibinfo {year}
  {2007})}\BibitemShut {NoStop}%
\bibitem [{\citenamefont {Wang}\ \emph {et~al.}(2010)\citenamefont {Wang},
  \citenamefont {Lv}, \citenamefont {Zhu},\ and\ \citenamefont
  {Ma}}]{CALYPSO1}%
  \BibitemOpen
  \bibfield  {author} {\bibinfo {author} {\bibfnamefont {Y.}~\bibnamefont
  {Wang}}, \bibinfo {author} {\bibfnamefont {J.}~\bibnamefont {Lv}}, \bibinfo
  {author} {\bibfnamefont {L.}~\bibnamefont {Zhu}}, \ and\ \bibinfo {author}
  {\bibfnamefont {Y.}~\bibnamefont {Ma}},\ }\href@noop {} {\bibfield  {journal}
  {\bibinfo  {journal} {Phys. Rev. B}\ }\textbf {\bibinfo {volume} {82}},\
  \bibinfo {pages} {094116} (\bibinfo {year} {2010})}\BibitemShut {NoStop}%
\bibitem [{\citenamefont {Zhang}\ \emph {et~al.}(2017)\citenamefont {Zhang},
  \citenamefont {Wang}, \citenamefont {Wang}, \citenamefont {Zhang},\ and\
  \citenamefont {Ma}}]{CALYPSO2}%
  \BibitemOpen
  \bibfield  {author} {\bibinfo {author} {\bibfnamefont {Y.}~\bibnamefont
  {Zhang}}, \bibinfo {author} {\bibfnamefont {H.}~\bibnamefont {Wang}},
  \bibinfo {author} {\bibfnamefont {Y.}~\bibnamefont {Wang}}, \bibinfo {author}
  {\bibfnamefont {L.}~\bibnamefont {Zhang}}, \ and\ \bibinfo {author}
  {\bibfnamefont {Y.}~\bibnamefont {Ma}},\ }\href@noop {} {\bibfield  {journal}
  {\bibinfo  {journal} {Phys. Rev. X}\ }\textbf {\bibinfo {volume} {7}},\
  \bibinfo {pages} {011017} (\bibinfo {year} {2017})}\BibitemShut {NoStop}%
\bibitem [{\citenamefont {Glass}\ \emph {et~al.}(2006)\citenamefont {Glass},
  \citenamefont {Oganov},\ and\ \citenamefont {Hansen}}]{uspex}%
  \BibitemOpen
  \bibfield  {author} {\bibinfo {author} {\bibfnamefont {C.~W.}\ \bibnamefont
  {Glass}}, \bibinfo {author} {\bibfnamefont {A.~R.}\ \bibnamefont {Oganov}}, \
  and\ \bibinfo {author} {\bibfnamefont {N.}~\bibnamefont {Hansen}},\
  }\href@noop {} {\bibfield  {journal} {\bibinfo  {journal} {Computer Physics
  Communications}\ }\textbf {\bibinfo {volume} {175}},\ \bibinfo {pages} {713 }
  (\bibinfo {year} {2006})}\BibitemShut {NoStop}%
\bibitem [{\citenamefont {Oganov}\ \emph {et~al.}(2011)\citenamefont {Oganov},
  \citenamefont {Lyakhov},\ and\ \citenamefont {Valle}}]{uspex_1}%
  \BibitemOpen
  \bibfield  {author} {\bibinfo {author} {\bibfnamefont {A.~R.}\ \bibnamefont
  {Oganov}}, \bibinfo {author} {\bibfnamefont {A.~O.}\ \bibnamefont {Lyakhov}},
  \ and\ \bibinfo {author} {\bibfnamefont {M.}~\bibnamefont {Valle}},\
  }\href@noop {} {\bibfield  {journal} {\bibinfo  {journal} {Accounts of
  Chemical Research}\ }\textbf {\bibinfo {volume} {44}},\ \bibinfo {pages}
  {227} (\bibinfo {year} {2011})}\BibitemShut {NoStop}%
\bibitem [{usp(2013)}]{uspex_2}%
  \BibitemOpen
  \href@noop {} {\bibfield  {journal} {\bibinfo  {journal} {Computer Physics
  Communications}\ }\textbf {\bibinfo {volume} {184}},\ \bibinfo {pages} {1172
  } (\bibinfo {year} {2013})}\BibitemShut {NoStop}%
\bibitem [{\citenamefont {Lonie}\ and\ \citenamefont
  {Zurek}(2011)}]{LONIE2011372}%
  \BibitemOpen
  \bibfield  {author} {\bibinfo {author} {\bibfnamefont {D.~C.}\ \bibnamefont
  {Lonie}}\ and\ \bibinfo {author} {\bibfnamefont {E.}~\bibnamefont {Zurek}},\
  }\href {\doibase https://doi.org/10.1016/j.cpc.2010.07.048} {\bibfield
  {journal} {\bibinfo  {journal} {Computer Physics Communications}\ }\textbf
  {\bibinfo {volume} {182}},\ \bibinfo {pages} {372 } (\bibinfo {year}
  {2011})}\BibitemShut {NoStop}%
\bibitem [{\citenamefont {Noh}\ \emph {et~al.}(2019)\citenamefont {Noh},
  \citenamefont {Kim}, \citenamefont {Stein}, \citenamefont
  {Sanchez-Lengeling}, \citenamefont {Gregoire}, \citenamefont {Aspuru-Guzik},\
  and\ \citenamefont {Jung}}]{NOH2019}%
  \BibitemOpen
  \bibfield  {author} {\bibinfo {author} {\bibfnamefont {J.}~\bibnamefont
  {Noh}}, \bibinfo {author} {\bibfnamefont {J.}~\bibnamefont {Kim}}, \bibinfo
  {author} {\bibfnamefont {H.~S.}\ \bibnamefont {Stein}}, \bibinfo {author}
  {\bibfnamefont {B.}~\bibnamefont {Sanchez-Lengeling}}, \bibinfo {author}
  {\bibfnamefont {J.~M.}\ \bibnamefont {Gregoire}}, \bibinfo {author}
  {\bibfnamefont {A.}~\bibnamefont {Aspuru-Guzik}}, \ and\ \bibinfo {author}
  {\bibfnamefont {Y.}~\bibnamefont {Jung}},\ }\href {\doibase
  https://doi.org/10.1016/j.matt.2019.08.017} {\bibfield  {journal} {\bibinfo
  {journal} {Matter}\ } (\bibinfo {year} {2019}),\
  https://doi.org/10.1016/j.matt.2019.08.017}\BibitemShut {NoStop}%
\bibitem [{\citenamefont {Ryan}\ \emph {et~al.}(2018)\citenamefont {Ryan},
  \citenamefont {Lengyel},\ and\ \citenamefont {Shatruk}}]{Ryan18}%
  \BibitemOpen
  \bibfield  {author} {\bibinfo {author} {\bibfnamefont {K.}~\bibnamefont
  {Ryan}}, \bibinfo {author} {\bibfnamefont {J.}~\bibnamefont {Lengyel}}, \
  and\ \bibinfo {author} {\bibfnamefont {M.}~\bibnamefont {Shatruk}},\ }\href
  {\doibase 10.1021/jacs.8b03913} {\bibfield  {journal} {\bibinfo  {journal}
  {Journal of the American Chemical Society}\ }\textbf {\bibinfo {volume}
  {140}},\ \bibinfo {pages} {10158} (\bibinfo {year} {2018})}\BibitemShut
  {NoStop}%
\bibitem [{\citenamefont {Saal}\ \emph
  {et~al.}(2013{\natexlab{b}})\citenamefont {Saal}, \citenamefont {Kirklin},
  \citenamefont {Aykol}, \citenamefont {Meredig},\ and\ \citenamefont
  {Wolverton}}]{OQMD}%
  \BibitemOpen
  \bibfield  {author} {\bibinfo {author} {\bibfnamefont {J.~E.}\ \bibnamefont
  {Saal}}, \bibinfo {author} {\bibfnamefont {S.}~\bibnamefont {Kirklin}},
  \bibinfo {author} {\bibfnamefont {M.}~\bibnamefont {Aykol}}, \bibinfo
  {author} {\bibfnamefont {B.}~\bibnamefont {Meredig}}, \ and\ \bibinfo
  {author} {\bibfnamefont {C.}~\bibnamefont {Wolverton}},\ }\href@noop {}
  {\bibfield  {journal} {\bibinfo  {journal} {JOM}\ }\textbf {\bibinfo {volume}
  {65}},\ \bibinfo {pages} {1501} (\bibinfo {year}
  {2013}{\natexlab{b}})}\BibitemShut {NoStop}%
\bibitem [{\citenamefont {Jeitschko}(2000)}]{Jeitschko00}%
  \BibitemOpen
  \bibfield  {author} {\bibinfo {author} {\bibfnamefont {H.~R. K. T. H. K.
  L.~A.}\ \bibnamefont {Jeitschko}, \bibfnamefont {W.}},\ }\href
  {http://oqmd.org/materials/entry/15968} {\bibfield  {journal} {\bibinfo
  {journal} {Journal of Solid State Chemistry}\ }\textbf {\bibinfo {volume}
  {154}} (\bibinfo {year} {2000})}\BibitemShut {NoStop}%
\bibitem [{\citenamefont {Kuzma}(1972)}]{Kuzma72}%
  \BibitemOpen
  \bibfield  {author} {\bibinfo {author} {\bibfnamefont {S.~S.}\ \bibnamefont
  {Kuzma}, \bibfnamefont {Yu.b.}},\ }\href
  {http://oqmd.org/materials/entry/3348} {\bibfield  {journal} {\bibinfo
  {journal} {Kristallografiya}\ }\textbf {\bibinfo {volume} {17}} (\bibinfo
  {year} {1972})}\BibitemShut {NoStop}%
\bibitem [{\citenamefont {Geupel}\ \emph {et~al.}(2001)\citenamefont {Geupel},
  \citenamefont {Zahn}, \citenamefont {Paufler},\ and\ \citenamefont
  {Graw}}]{Geupel01}%
  \BibitemOpen
  \bibfield  {author} {\bibinfo {author} {\bibfnamefont {S.}~\bibnamefont
  {Geupel}}, \bibinfo {author} {\bibfnamefont {G.}~\bibnamefont {Zahn}},
  \bibinfo {author} {\bibfnamefont {P.}~\bibnamefont {Paufler}}, \ and\
  \bibinfo {author} {\bibfnamefont {G.}~\bibnamefont {Graw}},\ }\href
  {http://oqmd.org/materials/entry/648870} {\bibfield  {journal} {\bibinfo
  {journal} {Zeitschrift fuer Kristallographie - New Crystal Structures}\
  }\textbf {\bibinfo {volume} {216}} (\bibinfo {year} {2001})}\BibitemShut
  {NoStop}%
\bibitem [{\citenamefont {Niihara}\ \emph {et~al.}(1987)\citenamefont
  {Niihara}, \citenamefont {Yajima},\ and\ \citenamefont
  {Shishido}}]{Niihara87}%
  \BibitemOpen
  \bibfield  {author} {\bibinfo {author} {\bibfnamefont {K.}~\bibnamefont
  {Niihara}}, \bibinfo {author} {\bibfnamefont {S.}~\bibnamefont {Yajima}}, \
  and\ \bibinfo {author} {\bibfnamefont {T.}~\bibnamefont {Shishido}},\ }\href
  {http://oqmd.org/materials/entry/659939} {\bibfield  {journal} {\bibinfo
  {journal} {Journal of the Less-Common Metals}\ }\textbf {\bibinfo {volume}
  {135}} (\bibinfo {year} {1987})}\BibitemShut {NoStop}%
\bibitem [{\citenamefont {Akselrud}\ \emph {et~al.}(1984)\citenamefont
  {Akselrud}, \citenamefont {Kuzma}, \citenamefont {Pecharskii},\ and\
  \citenamefont {Bilonizhko}}]{Akselrud84}%
  \BibitemOpen
  \bibfield  {author} {\bibinfo {author} {\bibfnamefont {L.}~\bibnamefont
  {Akselrud}}, \bibinfo {author} {\bibfnamefont {Y.}~\bibnamefont {Kuzma}},
  \bibinfo {author} {\bibfnamefont {V.}~\bibnamefont {Pecharskii}}, \ and\
  \bibinfo {author} {\bibfnamefont {N.}~\bibnamefont {Bilonizhko}},\ }\href
  {http://oqmd.org/materials/entry/28261} {\bibfield  {journal} {\bibinfo
  {journal} {Soviet Physics, Crystallography}\ }\textbf {\bibinfo {volume}
  {29}} (\bibinfo {year} {1984})}\BibitemShut {NoStop}%
\bibitem [{\citenamefont {Kuzma}\ \emph {et~al.}(1973)\citenamefont {Kuzma},
  \citenamefont {Svarichevskaya},\ and\ \citenamefont {Fomenko}}]{Kuzma73-1}%
  \BibitemOpen
  \bibfield  {author} {\bibinfo {author} {\bibfnamefont {Y.}~\bibnamefont
  {Kuzma}}, \bibinfo {author} {\bibfnamefont {S.}~\bibnamefont
  {Svarichevskaya}}, \ and\ \bibinfo {author} {\bibfnamefont {V.}~\bibnamefont
  {Fomenko}},\ }\href {http://oqmd.org/materials/entry/28255} {\bibfield
  {journal} {\bibinfo  {journal} {Inorganic Materials (USSR), see:
  Izv.Akad.Nauk, Neorg.Mater.}\ }\textbf {\bibinfo {volume} {9}} (\bibinfo
  {year} {1973})}\BibitemShut {NoStop}%
\bibitem [{\citenamefont {Jung}\ and\ \citenamefont
  {Schweitzer}(1986)}]{Jung86}%
  \BibitemOpen
  \bibfield  {author} {\bibinfo {author} {\bibfnamefont {W.}~\bibnamefont
  {Jung}}\ and\ \bibinfo {author} {\bibfnamefont {K.}~\bibnamefont
  {Schweitzer}},\ }\href {http://oqmd.org/materials/entry/11073} {\bibfield
  {journal} {\bibinfo  {journal} {Zeitschrift fuer Anorganische und Allgemeine
  Chemie}\ }\textbf {\bibinfo {volume} {533}} (\bibinfo {year}
  {1986})}\BibitemShut {NoStop}%
\bibitem [{\citenamefont {Jung}(1990{\natexlab{a}})}]{Jung90}%
  \BibitemOpen
  \bibfield  {author} {\bibinfo {author} {\bibfnamefont {W.}~\bibnamefont
  {Jung}},\ }\href {http://oqmd.org/materials/entry/12507} {\bibfield
  {journal} {\bibinfo  {journal} {Journal of the Less-Common Metals}\ }\textbf
  {\bibinfo {volume} {161}} (\bibinfo {year} {1990}{\natexlab{a}})}\BibitemShut
  {NoStop}%
\bibitem [{\citenamefont {Kuzma}\ and\ \citenamefont
  {Bilonizhko}(1973{\natexlab{a}})}]{Kuzma73-2}%
  \BibitemOpen
  \bibfield  {author} {\bibinfo {author} {\bibfnamefont {Y.}~\bibnamefont
  {Kuzma}}\ and\ \bibinfo {author} {\bibfnamefont {N.}~\bibnamefont
  {Bilonizhko}},\ }\href {http://oqmd.org/materials/entry/7681} {\bibfield
  {journal} {\bibinfo  {journal} {Kristallografiya}\ }\textbf {\bibinfo
  {volume} {18}} (\bibinfo {year} {1973}{\natexlab{a}})}\BibitemShut {NoStop}%
\bibitem [{\citenamefont {Akselrud}\ \emph {et~al.}(1985)\citenamefont
  {Akselrud}, \citenamefont {Kuzma},\ and\ \citenamefont
  {Bruskov}}]{Akselrud85}%
  \BibitemOpen
  \bibfield  {author} {\bibinfo {author} {\bibfnamefont {L.}~\bibnamefont
  {Akselrud}}, \bibinfo {author} {\bibfnamefont {Y.}~\bibnamefont {Kuzma}}, \
  and\ \bibinfo {author} {\bibfnamefont {V.}~\bibnamefont {Bruskov}},\ }\href
  {http://oqmd.org/materials/entry/56432} {\bibfield  {journal} {\bibinfo
  {journal} {Dopovidi Akademii Nauk Ukrainskoi RSR, Seriya B: Geologichni,
  Khimichni ta Biologichni Nauki}\ } (\bibinfo {year} {1985})}\BibitemShut
  {NoStop}%
\bibitem [{\citenamefont {Liang}\ \emph {et~al.}(2001)\citenamefont {Liang},
  \citenamefont {Rao}, \citenamefont {Chu},\ and\ \citenamefont
  {Yang}}]{Liang01}%
  \BibitemOpen
  \bibfield  {author} {\bibinfo {author} {\bibfnamefont {J.-k.}\ \bibnamefont
  {Liang}}, \bibinfo {author} {\bibfnamefont {G.-h.}\ \bibnamefont {Rao}},
  \bibinfo {author} {\bibfnamefont {W.}~\bibnamefont {Chu}}, \ and\ \bibinfo
  {author} {\bibfnamefont {L.~G.-y.}\ \bibnamefont {Yang}, \bibfnamefont
  {H.f.}},\ }\href {http://oqmd.org/materials/entry/16214} {\bibfield
  {journal} {\bibinfo  {journal} {Journal of Applied Physics}\ }\textbf
  {\bibinfo {volume} {90}} (\bibinfo {year} {2001})}\BibitemShut {NoStop}%
\bibitem [{\citenamefont {Liang}\ \emph {et~al.}(2000)\citenamefont {Liang},
  \citenamefont {Chen}, \citenamefont {Rao}, \citenamefont {Chen},
  \citenamefont {Liu}, \citenamefont {Shen}, \citenamefont {Li}, \citenamefont
  {Jin},\ and\ \citenamefont {Wang}}]{Liang00}%
  \BibitemOpen
  \bibfield  {author} {\bibinfo {author} {\bibfnamefont {J.-k.}\ \bibnamefont
  {Liang}}, \bibinfo {author} {\bibfnamefont {X.-l.}\ \bibnamefont {Chen}},
  \bibinfo {author} {\bibfnamefont {G.-h.}\ \bibnamefont {Rao}}, \bibinfo
  {author} {\bibfnamefont {Y.}~\bibnamefont {Chen}}, \bibinfo {author}
  {\bibfnamefont {Q.}~\bibnamefont {Liu}}, \bibinfo {author} {\bibfnamefont
  {B.-g.}\ \bibnamefont {Shen}}, \bibinfo {author} {\bibfnamefont
  {X.}~\bibnamefont {Li}}, \bibinfo {author} {\bibfnamefont {L.}~\bibnamefont
  {Jin}}, \ and\ \bibinfo {author} {\bibfnamefont {M.}~\bibnamefont {Wang}},\
  }\href {http://oqmd.org/materials/entry/15843} {\bibfield  {journal}
  {\bibinfo  {journal} {Chemistry of Materials}\ }\textbf {\bibinfo {volume}
  {12}} (\bibinfo {year} {2000})}\BibitemShut {NoStop}%
\bibitem [{\citenamefont {Noel}\ \emph {et~al.}(2003)\citenamefont {Noel},
  \citenamefont {Potel}, \citenamefont {Godart}, \citenamefont {O.l.},
  \citenamefont {Alleno},\ and\ \citenamefont {Salamakha}}]{Noel03}%
  \BibitemOpen
  \bibfield  {author} {\bibinfo {author} {\bibfnamefont {H.}~\bibnamefont
  {Noel}}, \bibinfo {author} {\bibfnamefont {M.}~\bibnamefont {Potel}},
  \bibinfo {author} {\bibfnamefont {S.}~\bibnamefont {Godart}, \bibfnamefont
  {C.}}, \bibinfo {author} {\bibfnamefont {C.}~\bibnamefont {O.l.},
  \bibfnamefont {Mazumdar}}, \bibinfo {author} {\bibfnamefont {E.}~\bibnamefont
  {Alleno}}, \ and\ \bibinfo {author} {\bibfnamefont {P.}~\bibnamefont
  {Salamakha}},\ }\href {http://oqmd.org/materials/entry/16547} {\bibfield
  {journal} {\bibinfo  {journal} {Journal of Alloys Compd.}\ }\textbf {\bibinfo
  {volume} {351}} (\bibinfo {year} {2003})}\BibitemShut {NoStop}%
\bibitem [{\citenamefont {Kuzma}(1981)}]{Kuzma81}%
  \BibitemOpen
  \bibfield  {author} {\bibinfo {author} {\bibfnamefont {B.~N.}\ \bibnamefont
  {Kuzma}, \bibfnamefont {Yu.b.}},\ }\href@noop {} {\bibfield  {journal}
  {\bibinfo  {journal} {Dopovidi Akademii Nauk Ukrainskoi RSR, Seriya A:
  Fiziko-Matematichni Ta Tekhnichni Nauki}\ }\textbf {\bibinfo {volume} {43}}
  (\bibinfo {year} {1981})}\BibitemShut {NoStop}%
\bibitem [{\citenamefont {Kuzma}\ and\ \citenamefont
  {Bilonizhko}(1973{\natexlab{b}})}]{Kuzma73-3}%
  \BibitemOpen
  \bibfield  {author} {\bibinfo {author} {\bibfnamefont {Y.}~\bibnamefont
  {Kuzma}}\ and\ \bibinfo {author} {\bibfnamefont {N.}~\bibnamefont
  {Bilonizhko}},\ }\href {http://oqmd.org/materials/entry/659890} {\bibfield
  {journal} {\bibinfo  {journal} {Soviet Physics, Crystallography}\ }\textbf
  {\bibinfo {volume} {18}} (\bibinfo {year} {1973}{\natexlab{b}})}\BibitemShut
  {NoStop}%
\bibitem [{\citenamefont {Kuzma}\ \emph {et~al.}(1989)\citenamefont {Kuzma},
  \citenamefont {Mikhalenko},\ and\ \citenamefont {Chaban}}]{Kuzma89}%
  \BibitemOpen
  \bibfield  {author} {\bibinfo {author} {\bibfnamefont {Y.}~\bibnamefont
  {Kuzma}}, \bibinfo {author} {\bibfnamefont {S.}~\bibnamefont {Mikhalenko}}, \
  and\ \bibinfo {author} {\bibfnamefont {N.}~\bibnamefont {Chaban}},\ }\href
  {http://oqmd.org/materials/entry/28263} {\bibfield  {journal} {\bibinfo
  {journal} {Soviet powder metallurgy and metal ceramics}\ }\textbf {\bibinfo
  {volume} {28}} (\bibinfo {year} {1989})}\BibitemShut {NoStop}%
\bibitem [{\citenamefont {Poettgen}\ \emph {et~al.}(2010)\citenamefont
  {Poettgen}, \citenamefont {Matar},\ and\ \citenamefont {M.}}]{Poettgen10}%
  \BibitemOpen
  \bibfield  {author} {\bibinfo {author} {\bibfnamefont {R.}~\bibnamefont
  {Poettgen}}, \bibinfo {author} {\bibfnamefont {E.}~\bibnamefont {Matar},
  \bibfnamefont {S.f.}}, \ and\ \bibinfo {author} {\bibfnamefont
  {T.}~\bibnamefont {M.}, \bibfnamefont {Mishra}},\ }\href
  {http://oqmd.org/materials/entry/654709} {\bibfield  {journal} {\bibinfo
  {journal} {Zeitschrift fuer Anorganische und Allgemeine Chemie}\ }\textbf
  {\bibinfo {volume} {636}} (\bibinfo {year} {2010})}\BibitemShut {NoStop}%
\bibitem [{\citenamefont {Jung}(1991)}]{Jung91}%
  \BibitemOpen
  \bibfield  {author} {\bibinfo {author} {\bibfnamefont {W.}~\bibnamefont
  {Jung}},\ }\href {http://oqmd.org/materials/entry/7315} {\bibfield  {journal}
  {\bibinfo  {journal} {Journal of the Less-Common Metals}\ }\textbf {\bibinfo
  {volume} {171}} (\bibinfo {year} {1991})}\BibitemShut {NoStop}%
\bibitem [{\citenamefont {Jung}(1990{\natexlab{b}})}]{Jung90-2}%
  \BibitemOpen
  \bibfield  {author} {\bibinfo {author} {\bibfnamefont {W.}~\bibnamefont
  {Jung}},\ }\href {http://oqmd.org/materials/entry/28387} {\bibfield
  {journal} {\bibinfo  {journal} {Journal of the Less-Common Metals}\ }\textbf
  {\bibinfo {volume} {161}} (\bibinfo {year} {1990}{\natexlab{b}})}\BibitemShut
  {NoStop}%
\bibitem [{\citenamefont {Kuzma}\ and\ \citenamefont
  {Bilonizhko}(1974)}]{Kuzma74}%
  \BibitemOpen
  \bibfield  {author} {\bibinfo {author} {\bibfnamefont {Y.}~\bibnamefont
  {Kuzma}}\ and\ \bibinfo {author} {\bibfnamefont {N.}~\bibnamefont
  {Bilonizhko}},\ }\href {http://oqmd.org/materials/entry/3800} {\bibfield
  {journal} {\bibinfo  {journal} {Izvestiya Akademii Nauk SSSR, Neorganicheskie
  Materialy}\ }\textbf {\bibinfo {volume} {10}} (\bibinfo {year}
  {1974})}\BibitemShut {NoStop}%
\bibitem [{\citenamefont {Kohn}\ and\ \citenamefont {Sham}(1965)}]{KS-DFT}%
  \BibitemOpen
  \bibfield  {author} {\bibinfo {author} {\bibfnamefont {W.}~\bibnamefont
  {Kohn}}\ and\ \bibinfo {author} {\bibfnamefont {L.~J.}\ \bibnamefont
  {Sham}},\ }\href@noop {} {\bibfield  {journal} {\bibinfo  {journal} {Phys.
  Rev.}\ }\textbf {\bibinfo {volume} {140}},\ \bibinfo {pages} {A1133}
  (\bibinfo {year} {1965})}\BibitemShut {NoStop}%
\bibitem [{\citenamefont {Hohenberg}\ and\ \citenamefont
  {Kohn}(1964)}]{HK-DFT}%
  \BibitemOpen
  \bibfield  {author} {\bibinfo {author} {\bibfnamefont {P.}~\bibnamefont
  {Hohenberg}}\ and\ \bibinfo {author} {\bibfnamefont {W.}~\bibnamefont
  {Kohn}},\ }\href@noop {} {\bibfield  {journal} {\bibinfo  {journal} {Phys.
  Rev.}\ }\textbf {\bibinfo {volume} {136}},\ \bibinfo {pages} {B864} (\bibinfo
  {year} {1964})}\BibitemShut {NoStop}%
\bibitem [{\citenamefont {Kirklin}\ \emph {et~al.}(2015)\citenamefont
  {Kirklin}, \citenamefont {Saal}, \citenamefont {Meredig}, \citenamefont
  {Thompson}, \citenamefont {Doak}, \citenamefont {Aykol}, \citenamefont
  {R{\"u}hl},\ and\ \citenamefont {Wolverton}}]{OQMD1}%
  \BibitemOpen
  \bibfield  {author} {\bibinfo {author} {\bibfnamefont {S.}~\bibnamefont
  {Kirklin}}, \bibinfo {author} {\bibfnamefont {J.~E.}\ \bibnamefont {Saal}},
  \bibinfo {author} {\bibfnamefont {B.}~\bibnamefont {Meredig}}, \bibinfo
  {author} {\bibfnamefont {A.}~\bibnamefont {Thompson}}, \bibinfo {author}
  {\bibfnamefont {J.~W.}\ \bibnamefont {Doak}}, \bibinfo {author}
  {\bibfnamefont {M.}~\bibnamefont {Aykol}}, \bibinfo {author} {\bibfnamefont
  {S.}~\bibnamefont {R{\"u}hl}}, \ and\ \bibinfo {author} {\bibfnamefont
  {C.}~\bibnamefont {Wolverton}},\ }\href@noop {} {\bibfield  {journal}
  {\bibinfo  {journal} {Npj Computational Materials}\ }\textbf {\bibinfo
  {volume} {1}},\ \bibinfo {pages} {15010 EP } (\bibinfo {year}
  {2015})}\BibitemShut {NoStop}%
\bibitem [{\citenamefont {Barber}\ \emph {et~al.}(1996)\citenamefont {Barber},
  \citenamefont {Dobkin},\ and\ \citenamefont {Huhdanpaa}}]{Barber96}%
  \BibitemOpen
  \bibfield  {author} {\bibinfo {author} {\bibfnamefont {C.~B.}\ \bibnamefont
  {Barber}}, \bibinfo {author} {\bibfnamefont {D.~P.}\ \bibnamefont {Dobkin}},
  \ and\ \bibinfo {author} {\bibfnamefont {H.}~\bibnamefont {Huhdanpaa}},\
  }\href {\doibase 10.1145/235815.235821} {\bibfield  {journal} {\bibinfo
  {journal} {ACM Trans. Math. Softw.}\ }\textbf {\bibinfo {volume} {22}},\
  \bibinfo {pages} {469–483} (\bibinfo {year} {1996})}\BibitemShut {NoStop}%
\bibitem [{\citenamefont {Balachandran}\ \emph {et~al.}(2018)\citenamefont
  {Balachandran}, \citenamefont {Emery}, \citenamefont {Gubernatis},
  \citenamefont {Lookman}, \citenamefont {Wolverton},\ and\ \citenamefont
  {Zunger}}]{AB_perovskite}%
  \BibitemOpen
  \bibfield  {author} {\bibinfo {author} {\bibfnamefont {P.~V.}\ \bibnamefont
  {Balachandran}}, \bibinfo {author} {\bibfnamefont {A.~A.}\ \bibnamefont
  {Emery}}, \bibinfo {author} {\bibfnamefont {J.~E.}\ \bibnamefont
  {Gubernatis}}, \bibinfo {author} {\bibfnamefont {T.}~\bibnamefont {Lookman}},
  \bibinfo {author} {\bibfnamefont {C.}~\bibnamefont {Wolverton}}, \ and\
  \bibinfo {author} {\bibfnamefont {A.}~\bibnamefont {Zunger}},\ }\href@noop {}
  {\bibfield  {journal} {\bibinfo  {journal} {Phys. Rev. Materials}\ }\textbf
  {\bibinfo {volume} {2}},\ \bibinfo {pages} {043802} (\bibinfo {year}
  {2018})}\BibitemShut {NoStop}%
\bibitem [{\citenamefont {Kresse}\ and\ \citenamefont {Hafner}(1993)}]{vasp1}%
  \BibitemOpen
  \bibfield  {author} {\bibinfo {author} {\bibfnamefont {G.}~\bibnamefont
  {Kresse}}\ and\ \bibinfo {author} {\bibfnamefont {J.}~\bibnamefont
  {Hafner}},\ }\href@noop {} {\bibfield  {journal} {\bibinfo  {journal} {Phys.
  Rev. B}\ }\textbf {\bibinfo {volume} {47}},\ \bibinfo {pages} {558} (\bibinfo
  {year} {1993})}\BibitemShut {NoStop}%
\bibitem [{\citenamefont {Kresse}\ and\ \citenamefont {Hafner}(1994)}]{vasp2}%
  \BibitemOpen
  \bibfield  {author} {\bibinfo {author} {\bibfnamefont {G.}~\bibnamefont
  {Kresse}}\ and\ \bibinfo {author} {\bibfnamefont {J.}~\bibnamefont
  {Hafner}},\ }\href@noop {} {\bibfield  {journal} {\bibinfo  {journal} {Phys.
  Rev. B}\ }\textbf {\bibinfo {volume} {49}},\ \bibinfo {pages} {14251}
  (\bibinfo {year} {1994})}\BibitemShut {NoStop}%
\bibitem [{\citenamefont {Kresse}\ and\ \citenamefont
  {Furthmuller}(1996{\natexlab{a}})}]{vasp3}%
  \BibitemOpen
  \bibfield  {author} {\bibinfo {author} {\bibfnamefont {G.}~\bibnamefont
  {Kresse}}\ and\ \bibinfo {author} {\bibfnamefont {J.}~\bibnamefont
  {Furthmuller}},\ }\href@noop {} {\bibfield  {journal} {\bibinfo  {journal}
  {Comput. Mat. Sci.}\ }\textbf {\bibinfo {volume} {6}},\ \bibinfo {pages} {15
  } (\bibinfo {year} {1996}{\natexlab{a}})}\BibitemShut {NoStop}%
\bibitem [{\citenamefont {Kresse}\ and\ \citenamefont
  {Furthmuller}(1996{\natexlab{b}})}]{vasp4}%
  \BibitemOpen
  \bibfield  {author} {\bibinfo {author} {\bibfnamefont {G.}~\bibnamefont
  {Kresse}}\ and\ \bibinfo {author} {\bibfnamefont {J.}~\bibnamefont
  {Furthmuller}},\ }\href@noop {} {\bibfield  {journal} {\bibinfo  {journal}
  {Phys. Rev. B}\ }\textbf {\bibinfo {volume} {54}},\ \bibinfo {pages} {11169}
  (\bibinfo {year} {1996}{\natexlab{b}})}\BibitemShut {NoStop}%
\bibitem [{\citenamefont {Bl\"ochl}(1994)}]{paw1}%
  \BibitemOpen
  \bibfield  {author} {\bibinfo {author} {\bibfnamefont {P.~E.}\ \bibnamefont
  {Bl\"ochl}},\ }\href@noop {} {\bibfield  {journal} {\bibinfo  {journal}
  {Phys. Rev. B}\ }\textbf {\bibinfo {volume} {50}},\ \bibinfo {pages} {17953}
  (\bibinfo {year} {1994})}\BibitemShut {NoStop}%
\bibitem [{\citenamefont {Kresse}\ and\ \citenamefont {Joubert}(1999)}]{paw2}%
  \BibitemOpen
  \bibfield  {author} {\bibinfo {author} {\bibfnamefont {G.}~\bibnamefont
  {Kresse}}\ and\ \bibinfo {author} {\bibfnamefont {D.}~\bibnamefont
  {Joubert}},\ }\href@noop {} {\bibfield  {journal} {\bibinfo  {journal} {Phys.
  Rev. B}\ }\textbf {\bibinfo {volume} {59}},\ \bibinfo {pages} {1758}
  (\bibinfo {year} {1999})}\BibitemShut {NoStop}%
\bibitem [{\citenamefont {Perdew}\ \emph {et~al.}(1996)\citenamefont {Perdew},
  \citenamefont {Burke},\ and\ \citenamefont {Ernzerhof}}]{pbe}%
  \BibitemOpen
  \bibfield  {author} {\bibinfo {author} {\bibfnamefont {J.~P.}\ \bibnamefont
  {Perdew}}, \bibinfo {author} {\bibfnamefont {K.}~\bibnamefont {Burke}}, \
  and\ \bibinfo {author} {\bibfnamefont {M.}~\bibnamefont {Ernzerhof}},\
  }\href@noop {} {\bibfield  {journal} {\bibinfo  {journal} {Phys. Rev. Lett}\
  }\textbf {\bibinfo {volume} {77}},\ \bibinfo {pages} {3865} (\bibinfo {year}
  {1996})}\BibitemShut {NoStop}%
\bibitem [{\citenamefont {Lam~Pham}\ \emph {et~al.}(2017)\citenamefont
  {Lam~Pham}, \citenamefont {Kino}, \citenamefont {Terakura}, \citenamefont
  {Miyake}, \citenamefont {Tsuda}, \citenamefont {Takigawa},\ and\
  \citenamefont {Chi~Dam}}]{ofm}%
  \BibitemOpen
  \bibfield  {author} {\bibinfo {author} {\bibfnamefont {T.}~\bibnamefont
  {Lam~Pham}}, \bibinfo {author} {\bibfnamefont {H.}~\bibnamefont {Kino}},
  \bibinfo {author} {\bibfnamefont {K.}~\bibnamefont {Terakura}}, \bibinfo
  {author} {\bibfnamefont {T.}~\bibnamefont {Miyake}}, \bibinfo {author}
  {\bibfnamefont {K.}~\bibnamefont {Tsuda}}, \bibinfo {author} {\bibfnamefont
  {I.}~\bibnamefont {Takigawa}}, \ and\ \bibinfo {author} {\bibfnamefont
  {H.}~\bibnamefont {Chi~Dam}},\ }\href@noop {} {\bibfield  {journal} {\bibinfo
   {journal} {Sci Technol Adv Mater}\ }\textbf {\bibinfo {volume} {18}},\
  \bibinfo {pages} {756} (\bibinfo {year} {2017})}\BibitemShut {NoStop}%
\bibitem [{\citenamefont {Pham}\ \emph {et~al.}(2018)\citenamefont {Pham},
  \citenamefont {Nguyen}, \citenamefont {Nguyen}, \citenamefont {Kino},
  \citenamefont {Miyake},\ and\ \citenamefont {Dam}}]{ofm1}%
  \BibitemOpen
  \bibfield  {author} {\bibinfo {author} {\bibfnamefont {T.-L.}\ \bibnamefont
  {Pham}}, \bibinfo {author} {\bibfnamefont {N.-D.}\ \bibnamefont {Nguyen}},
  \bibinfo {author} {\bibfnamefont {V.-D.}\ \bibnamefont {Nguyen}}, \bibinfo
  {author} {\bibfnamefont {H.}~\bibnamefont {Kino}}, \bibinfo {author}
  {\bibfnamefont {T.}~\bibnamefont {Miyake}}, \ and\ \bibinfo {author}
  {\bibfnamefont {H.-C.}\ \bibnamefont {Dam}},\ }\href@noop {} {\bibfield
  {journal} {\bibinfo  {journal} {The Journal of Chemical Physics}\ }\textbf
  {\bibinfo {volume} {148}},\ \bibinfo {pages} {204106} (\bibinfo {year}
  {2018})}\BibitemShut {NoStop}%
\bibitem [{\citenamefont {Murphy}(2012)}]{ML}%
  \BibitemOpen
  \bibfield  {author} {\bibinfo {author} {\bibfnamefont {K.~P.}\ \bibnamefont
  {Murphy}},\ }\enquote {\bibinfo {title} {Machine learning: A probabilistic
  perspective},}\ \ (\bibinfo  {publisher} {MIT Press},\ \bibinfo {year}
  {2012})\BibitemShut {NoStop}%
\bibitem [{\citenamefont {Kvalseth}(1985)}]{Kvalseth85}%
  \BibitemOpen
  \bibfield  {author} {\bibinfo {author} {\bibfnamefont {T.~O.}\ \bibnamefont
  {Kvalseth}},\ }\href@noop {} {\bibfield  {journal} {\bibinfo  {journal} {The
  American Statistician}\ }\textbf {\bibinfo {volume} {39}} (\bibinfo {year}
  {1985})}\BibitemShut {NoStop}%
\bibitem [{\citenamefont {Nguyen}\ \emph {et~al.}(2019)\citenamefont {Nguyen},
  \citenamefont {Pham}, \citenamefont {Nguyen}, \citenamefont {Nguyen},
  \citenamefont {Kino}, \citenamefont {Miyake},\ and\ \citenamefont
  {Dam}}]{Nguyen_2019}%
  \BibitemOpen
  \bibfield  {author} {\bibinfo {author} {\bibfnamefont {D.-N.}\ \bibnamefont
  {Nguyen}}, \bibinfo {author} {\bibfnamefont {T.-L.}\ \bibnamefont {Pham}},
  \bibinfo {author} {\bibfnamefont {V.-C.}\ \bibnamefont {Nguyen}}, \bibinfo
  {author} {\bibfnamefont {A.-T.}\ \bibnamefont {Nguyen}}, \bibinfo {author}
  {\bibfnamefont {H.}~\bibnamefont {Kino}}, \bibinfo {author} {\bibfnamefont
  {T.}~\bibnamefont {Miyake}}, \ and\ \bibinfo {author} {\bibfnamefont {H.-C.}\
  \bibnamefont {Dam}},\ }\href {\doibase 10.1088/1742-6596/1290/1/012009}
  {\bibfield  {journal} {\bibinfo  {journal} {Journal of Physics: Conference
  Series}\ }\textbf {\bibinfo {volume} {1290}},\ \bibinfo {pages} {012009}
  (\bibinfo {year} {2019})}\BibitemShut {NoStop}%
\bibitem [{\citenamefont {Yu}\ and\ \citenamefont
  {Liu}(2004)}]{feature_selection_Yu:2004}%
  \BibitemOpen
  \bibfield  {author} {\bibinfo {author} {\bibfnamefont {L.}~\bibnamefont
  {Yu}}\ and\ \bibinfo {author} {\bibfnamefont {H.}~\bibnamefont {Liu}},\
  }\href@noop {} {\bibfield  {journal} {\bibinfo  {journal} {J. Mach. Learn.
  Res.}\ }\textbf {\bibinfo {volume} {5}},\ \bibinfo {pages} {1205} (\bibinfo
  {year} {2004})}\BibitemShut {NoStop}%
\bibitem [{\citenamefont {Visalakshi}\ and\ \citenamefont
  {Radha}(2014)}]{feature_selection}%
  \BibitemOpen
  \bibfield  {author} {\bibinfo {author} {\bibfnamefont {S.}~\bibnamefont
  {Visalakshi}}\ and\ \bibinfo {author} {\bibfnamefont {V.}~\bibnamefont
  {Radha}}\ }(\bibinfo {year} {2014})\ pp.\ \bibinfo {pages} {1--6}\BibitemShut
  {NoStop}%
\bibitem [{\citenamefont {Nguyen}\ \emph {et~al.}(2018)\citenamefont {Nguyen},
  \citenamefont {Pham}, \citenamefont {Nguyen}, \citenamefont {Ho},
  \citenamefont {Tran}, \citenamefont {Takahashi},\ and\ \citenamefont
  {Dam}}]{Nguyen18}%
  \BibitemOpen
  \bibfield  {author} {\bibinfo {author} {\bibfnamefont {D.-N.}\ \bibnamefont
  {Nguyen}}, \bibinfo {author} {\bibfnamefont {T.-L.}\ \bibnamefont {Pham}},
  \bibinfo {author} {\bibfnamefont {V.-C.}\ \bibnamefont {Nguyen}}, \bibinfo
  {author} {\bibfnamefont {T.-D.}\ \bibnamefont {Ho}}, \bibinfo {author}
  {\bibfnamefont {T.}~\bibnamefont {Tran}}, \bibinfo {author} {\bibfnamefont
  {K.}~\bibnamefont {Takahashi}}, \ and\ \bibinfo {author} {\bibfnamefont
  {H.-C.}\ \bibnamefont {Dam}},\ }\href {\doibase 10.1107/S2052252518013519}
  {\bibfield  {journal} {\bibinfo  {journal} {IUCrJ}\ }\textbf {\bibinfo
  {volume} {5}},\ \bibinfo {pages} {830} (\bibinfo {year} {2018})}\BibitemShut
  {NoStop}%
\bibitem [{\citenamefont {Dam}\ \emph {et~al.}(2018)\citenamefont {Dam},
  \citenamefont {Nguyen}, \citenamefont {Pham}, \citenamefont {Nguyen},
  \citenamefont {Terakura}, \citenamefont {Miyake},\ and\ \citenamefont
  {Kino}}]{Dam18}%
  \BibitemOpen
  \bibfield  {author} {\bibinfo {author} {\bibfnamefont {H.~C.}\ \bibnamefont
  {Dam}}, \bibinfo {author} {\bibfnamefont {V.~C.}\ \bibnamefont {Nguyen}},
  \bibinfo {author} {\bibfnamefont {T.~L.}\ \bibnamefont {Pham}}, \bibinfo
  {author} {\bibfnamefont {A.~T.}\ \bibnamefont {Nguyen}}, \bibinfo {author}
  {\bibfnamefont {K.}~\bibnamefont {Terakura}}, \bibinfo {author}
  {\bibfnamefont {T.}~\bibnamefont {Miyake}}, \ and\ \bibinfo {author}
  {\bibfnamefont {H.}~\bibnamefont {Kino}},\ }\href {\doibase
  10.7566/JPSJ.87.113801} {\bibfield  {journal} {\bibinfo  {journal} {Journal
  of the Physical Society of Japan}\ }\textbf {\bibinfo {volume} {87}},\
  \bibinfo {pages} {113801} (\bibinfo {year} {2018})},\ \Eprint
  {http://arxiv.org/abs/https://doi.org/10.7566/JPSJ.87.113801}
  {https://doi.org/10.7566/JPSJ.87.113801} \BibitemShut {NoStop}%
\bibitem [{\citenamefont {Pedregosa}\ \emph {et~al.}(2011)\citenamefont
  {Pedregosa}, \citenamefont {Varoquaux}, \citenamefont {Gramfort},
  \citenamefont {Michel}, \citenamefont {Thirion}, \citenamefont {Grisel},
  \citenamefont {Blondel}, \citenamefont {Prettenhofer}, \citenamefont {Weiss},
  \citenamefont {Dubourg}, \citenamefont {Vanderplas}, \citenamefont {Passos},
  \citenamefont {Cournapeau}, \citenamefont {Brucher}, \citenamefont {Perrot},\
  and\ \citenamefont {Duchesnay}}]{scikit-learn}%
  \BibitemOpen
  \bibfield  {author} {\bibinfo {author} {\bibfnamefont {F.}~\bibnamefont
  {Pedregosa}}, \bibinfo {author} {\bibfnamefont {G.}~\bibnamefont
  {Varoquaux}}, \bibinfo {author} {\bibfnamefont {A.}~\bibnamefont {Gramfort}},
  \bibinfo {author} {\bibfnamefont {V.}~\bibnamefont {Michel}}, \bibinfo
  {author} {\bibfnamefont {B.}~\bibnamefont {Thirion}}, \bibinfo {author}
  {\bibfnamefont {O.}~\bibnamefont {Grisel}}, \bibinfo {author} {\bibfnamefont
  {M.}~\bibnamefont {Blondel}}, \bibinfo {author} {\bibfnamefont
  {P.}~\bibnamefont {Prettenhofer}}, \bibinfo {author} {\bibfnamefont
  {R.}~\bibnamefont {Weiss}}, \bibinfo {author} {\bibfnamefont
  {V.}~\bibnamefont {Dubourg}}, \bibinfo {author} {\bibfnamefont
  {J.}~\bibnamefont {Vanderplas}}, \bibinfo {author} {\bibfnamefont
  {A.}~\bibnamefont {Passos}}, \bibinfo {author} {\bibfnamefont
  {D.}~\bibnamefont {Cournapeau}}, \bibinfo {author} {\bibfnamefont
  {M.}~\bibnamefont {Brucher}}, \bibinfo {author} {\bibfnamefont
  {M.}~\bibnamefont {Perrot}}, \ and\ \bibinfo {author} {\bibfnamefont
  {E.}~\bibnamefont {Duchesnay}},\ }\href@noop {} {\bibfield  {journal}
  {\bibinfo  {journal} {J Mach Learn Res}\ }\textbf {\bibinfo {volume} {12}},\
  \bibinfo {pages} {2825} (\bibinfo {year} {2011})}\BibitemShut {NoStop}%
\bibitem [{\citenamefont {Ng}(2004)}]{ANg04}%
  \BibitemOpen
  \bibfield  {author} {\bibinfo {author} {\bibfnamefont {A.~Y.}\ \bibnamefont
  {Ng}},\ }\href@noop {} {\bibfield  {journal} {\bibinfo  {journal}
  {International Conference on Machine Learning}\ } (\bibinfo {year}
  {2004})}\BibitemShut {NoStop}%
\bibitem [{\citenamefont {Lee}\ \emph {et~al.}(2006)\citenamefont {Lee},
  \citenamefont {Lee}, \citenamefont {Abbeel},\ and\ \citenamefont
  {Y.~Ng}}]{Lee06}%
  \BibitemOpen
  \bibfield  {author} {\bibinfo {author} {\bibfnamefont {S.-I.}\ \bibnamefont
  {Lee}}, \bibinfo {author} {\bibfnamefont {H.}~\bibnamefont {Lee}}, \bibinfo
  {author} {\bibfnamefont {P.}~\bibnamefont {Abbeel}}, \ and\ \bibinfo {author}
  {\bibfnamefont {A.}~\bibnamefont {Y.~Ng}},\ }\href@noop {} {\bibfield
  {journal} {\bibinfo  {journal} {Proceedings of the 21st National Conference
  on Artificial Intelligence (AAAI-06)}\ }\textbf {\bibinfo {volume} {21}}
  (\bibinfo {year} {2006})}\BibitemShut {NoStop}%
\bibitem [{\citenamefont {Breiman}\ \emph {et~al.}(1984)\citenamefont
  {Breiman}, \citenamefont {Friedman}, \citenamefont {Stone},\ and\
  \citenamefont {Olshen}}]{Breiman84}%
  \BibitemOpen
  \bibfield  {author} {\bibinfo {author} {\bibfnamefont {L.}~\bibnamefont
  {Breiman}}, \bibinfo {author} {\bibfnamefont {J.}~\bibnamefont {Friedman}},
  \bibinfo {author} {\bibfnamefont {C.}~\bibnamefont {Stone}}, \ and\ \bibinfo
  {author} {\bibfnamefont {R.}~\bibnamefont {Olshen}},\ }\href@noop {}
  {\bibfield  {journal} {\bibinfo  {journal} {Wadsworth, Belmont}\ } (\bibinfo
  {year} {1984})}\BibitemShut {NoStop}%
\bibitem [{\citenamefont {Hastie}\ \emph {et~al.}(2009)\citenamefont {Hastie},
  \citenamefont {Tibshirani},\ and\ \citenamefont
  {Friedman}}]{hastie2009elements}%
  \BibitemOpen
  \bibfield  {author} {\bibinfo {author} {\bibfnamefont {T.}~\bibnamefont
  {Hastie}}, \bibinfo {author} {\bibfnamefont {R.}~\bibnamefont {Tibshirani}},
  \ and\ \bibinfo {author} {\bibfnamefont {J.~H.}\ \bibnamefont {Friedman}},\
  }\href {https://books.google.co.jp/books?id=eBSgoAEACAAJ} {\ \bibinfo
  {series} {Springer series in statistics} (\bibinfo {year}
  {2009})}\BibitemShut {NoStop}%
\bibitem [{\citenamefont {Perry}\ \emph {et~al.}(1955)\citenamefont {Perry},
  \citenamefont {Kent},\ and\ \citenamefont {Berry}}]{Perry55}%
  \BibitemOpen
  \bibfield  {author} {\bibinfo {author} {\bibfnamefont {J.~W.}\ \bibnamefont
  {Perry}}, \bibinfo {author} {\bibfnamefont {A.}~\bibnamefont {Kent}}, \ and\
  \bibinfo {author} {\bibfnamefont {M.~M.}\ \bibnamefont {Berry}},\ }\href
  {\doibase 10.1002/asi.5090060411} {\bibfield  {journal} {\bibinfo  {journal}
  {American Documentation}\ }\textbf {\bibinfo {volume} {6}},\ \bibinfo {pages}
  {242} (\bibinfo {year} {1955})},\ \Eprint
  {http://arxiv.org/abs/https://onlinelibrary.wiley.com/doi/pdf/10.1002/asi.5090060411}
  {https://onlinelibrary.wiley.com/doi/pdf/10.1002/asi.5090060411} \BibitemShut
  {NoStop}%
\bibitem [{\citenamefont {Derczynski}(2016)}]{Derczynski16}%
  \BibitemOpen
  \bibfield  {author} {\bibinfo {author} {\bibfnamefont {L.}~\bibnamefont
  {Derczynski}},\ }\href@noop {} {\bibfield  {journal} {\bibinfo  {journal}
  {Proceedings of the International Conference on Language Resources and
  Evaluation}\ } (\bibinfo {year} {2016})}\BibitemShut {NoStop}%
\bibitem [{\citenamefont {Su}\ \emph {et~al.}(2015)\citenamefont {Su},
  \citenamefont {Yuan},\ and\ \citenamefont {Zhu}}]{Su15}%
  \BibitemOpen
  \bibfield  {author} {\bibinfo {author} {\bibfnamefont {W.}~\bibnamefont
  {Su}}, \bibinfo {author} {\bibfnamefont {Y.}~\bibnamefont {Yuan}}, \ and\
  \bibinfo {author} {\bibfnamefont {M.}~\bibnamefont {Zhu}},\ }\href {\doibase
  10.1145/2808194.2809481} {\bibfield  {journal} {\bibinfo  {journal}
  {Proceedings of the 2015 International Conference on The Theory of
  Information Retrieval}\ }\bibinfo {series} {ICTIR '15},\ \bibinfo {pages}
  {349} (\bibinfo {year} {2015})}\BibitemShut {NoStop}%
\end{thebibliography}%

\section{Supplemental Materials}

\begin{table}[!th]
\centering
\caption{\label{tab:Nd-Fe-B}Ternary phases of Nd-Fe-B compounds: formation energy $E_f$ (eV/atom) and the stability calculated by DFT, $\Delta E^{DFT}$, (eV/atom) given by OQMD. }

\begin{tabular}{p{1.5cm}R{1.5cm}R{1.5cm}R{3.5cm}}
\hline\hline
Compound & \centering $E_f$ (eV/atom) & \centering $\Delta E^{DFT}$ (eV/atom) & Stability state\\
\hline
NdFe$_4$B$_4$  	&-0.432	  &  0.000   &  Potentially formable  \\
Nd$_5$Fe$_2$B$_6$ &-0.390  &  0.000  &   Potentially formable  \\
NdFe$_{12}$B$_6$ &-0.281   &  0.022   &   Potentially formable  \\
Nd$_4$Fe$_3$B$_6$	&-0.286	 &  0.118   &  Unstable \\
Nd$_2$FeB  	&0.446	 &  0.689  &   Unstable \\
NdFe$_2$B    	&0.775	 &  1.018   & Unstable  \\
NdFeB$_2$  	&0.714   	 &  1.145   &  Unstable  \\
\hline\hline
\end{tabular}
\end{table}

\begin{figure*}
	\includegraphics[scale=0.3]{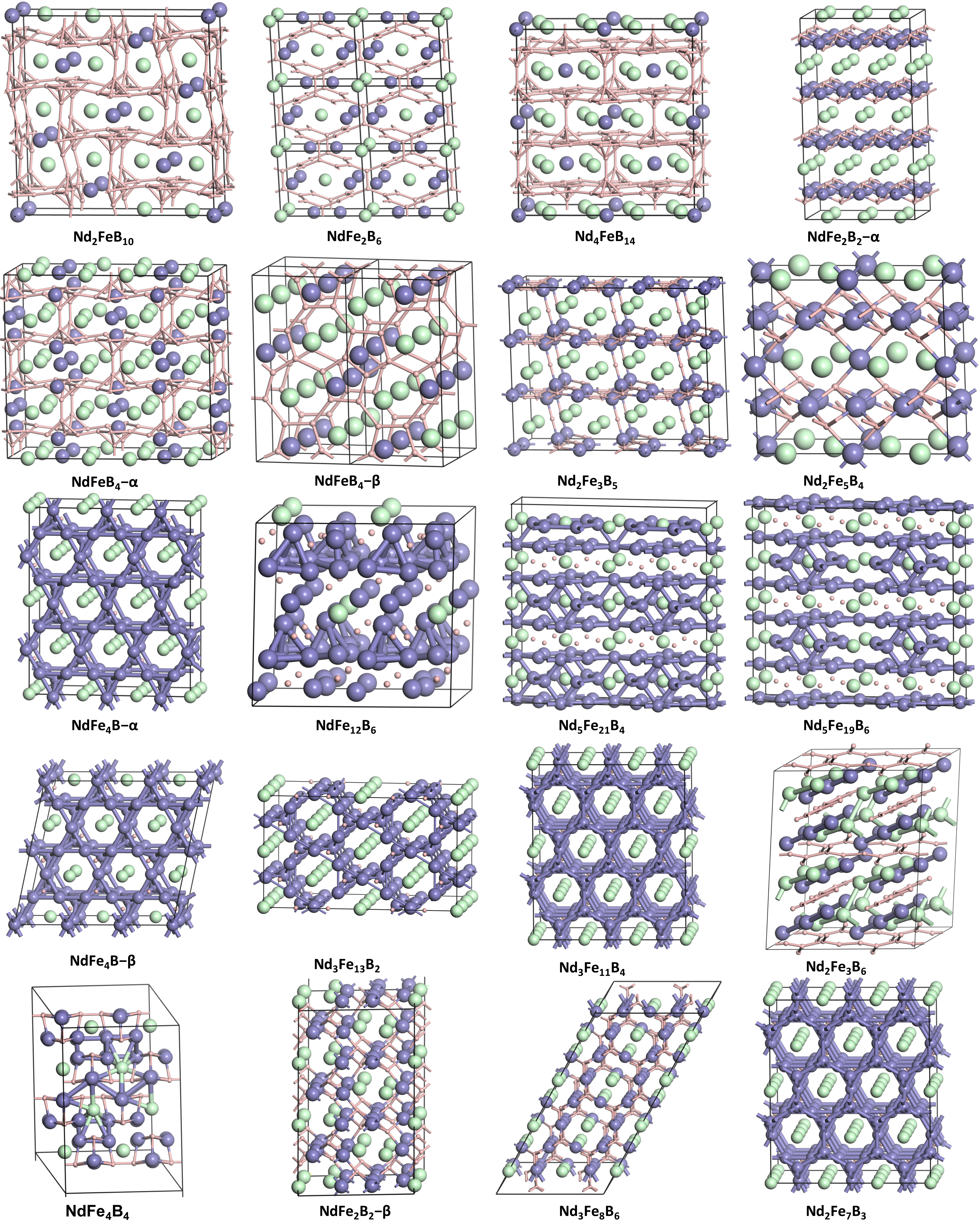}
    \caption{\label{fig:all_stab_struct}Twenty potentially formable Nd-Fe-B structures extracted by applying elemental substitution method to lanthanide--transition metal--light element materials.}
\end{figure*}

\end{document}